\documentclass[fleqn,usenatbib,twocolumn]{mnras}

\usepackage{newtxtext,newtxmath}
\usepackage[T1]{fontenc}
\usepackage{ae,aecompl}

\usepackage{graphicx}	% Including figure files
\usepackage{amsmath}	% Advanced maths commands
\usepackage{amssymb}	% Extra maths symbols
\usepackage{xcolor}

\title[X-AGN LSS mock]{Active Galactic Nuclei And Their Large-Scale Structure: An eROSITA Mock Catalogue}

\author[J. Comparat et al.]{
J. Comparat,$^{1}$\thanks{E-mail: comparat@mpe.mpg.de}
A. Merloni$^{1}$, 
M. Salvato$^{1}$,
K. Nandra$^{1}$,
T. Boller$^{1}$,
A. Georgakakis$^{2}$,
\newauthor  
A. Finoguenov$^{3}$,
T. Dwelly$^{1}$,
J. Buchner$^{4,5,6}$, 
A. Del Moro$^{1}$, 
N. Clerc$^{7}$,
Y. Wang$^{8}$, 
\newauthor
G. Zhao$^{8,9,10}$, 
F. Prada$^{11}$,
G. Yepes$^{12}$, 
M. Brusa$^{13,14}$,
M. Krumpe$^{15}$,
T. Liu$^{1}$
\\
$^{1}$ Max-Planck-Institut f\"{u}r extraterrestrische Physik (MPE), Giessenbachstrasse 1, D-85748 Garching bei M\"unchen, Germany \\
$^{2}$ Institute for Astronomy \& Astrophysics, National Observatory of Athens, V. Paulou \& I. Metaxa, 11532, Greece\\
$^{3}$ Department of Physics, University of Helsinki, Gustaf H\"allstr\"omin katu 2a, FI-00014 Helsinki, Finland\\
$^{4}$ Pontificia Universidad Cat\'olica de Chile, Instituto de Astrof\'isica, Casilla 306, Santiago 22, Chile\\
$^{5}$ Excellence Cluster Universe, Boltzmannstr. 2, D-85748, Garching, Germany\\
$^{6}$ Millenium Institute of Astrophysics, Vicu$\tilde{n}$a MacKenna 4860, 7820436 Macul, Santiago, Chile\\
$^{7}$ IRAP, Universit\'e de Toulouse, CNRS, UPS, CNES, 31400, Toulouse, France\\
$^{8}$ National Astronomy Observatories, Chinese Academy of Science, Beijing, 100101, P.R.China\\
$^{9}$ School of Astronomy and Space Science, University of Chinese Academy of Sciences, Beijing, 100049, P.R.China\\
$^{10}$ Institute of Cosmology and Gravitation, University of Portsmouth, Portsmouth, PO1 3FX, UK\\
$^{11}$ Instituto de Astrof\'{\i}sica de Andaluc\'{\i}a (CSIC), Glorieta de la Astronom\'{\i}a, E-18080 Granada, Spain \\
$^{12}$ Departamento de F\'{\i}sica Te\'orica and CIAFF, Universidad Aut\'onoma de Madrid, 28049 Madrid, Spain\\
$^{13}$ Dipartimento di Fisica e Astronomia, Alma Mater Studiorum Universita di Bologna, via Gobetti 93/2, 40129 Bologna, Italy \\
$^{14}$ INAF - Osservatorio di Astrofisica e Scienza dello Spazio di Bologna, via Gobetti 93/3, 40129 Bologna, Italy\\
$^{15}$ Leibniz-Institut f\"{u}r Astrophysik Potsdam, An der Sternwarte 16, 14482 Potsdam, Germany\\
}
\date{\today}
\pubyear{42}
\begin{document}
\label{firstpage}
\pagerange{\pageref{firstpage}--\pageref{lastpage}}
\maketitle
% Abstract of the paper
\begin{abstract}
In the context of the upcoming SRG/eROSITA survey, we present an N-body simulation-based mock catalogue for X-ray selected AGN samples. 
The model reproduces the observed hard X-ray AGN luminosity function (XLF) and the soft X-ray logN-logS from redshift 0 to 6.
The XLF is reproduced to within $\pm5\%$ and the logN-logS to within $\pm20\%$. 
We develop a joint X-ray -- optical extinction and classification model. 
We adopt a set of empirical spectral energy distributions to predict observed magnitudes in the UV, optical and NIR.
With the latest eROSITA all sky survey sensitivity model, we create a high-fidelity full-sky mock catalogue of X-ray AGN. 
It predicts their distributions in right ascension, declination, redshift and fluxes. 
Using empirical medium resolution optical spectral templates and an exposure time calculator, we find that $1.1\times10^6$ ($4\times10^5$) fiber-hours are needed to follow-up spectroscopically from the ground the detected X-ray AGN with an optical magnitude $21<r<22.8$ ($22.8<r<25$) with a 4-m (8-m) class multi-object spectroscopic facility.
We find that future clustering studies will measure the AGN bias to the percent level at redshift $z<1.2$ and should discriminate possible scenarios of galaxy-AGN co-evolution. 
We predict the accuracy to which the baryon acoustic oscillation standard ruler will be measured using X-ray
AGN: better than 3\% for AGN between redshift 0.5 to 3 and better than 1\% using the Ly$\alpha$ forest of X-ray QSOs discovered between redshift 2 and 3. 
eROSITA will provide an outstanding set of targets for future galaxy evolution and cosmological studies.
\end{abstract}


%%%%%%%%%%%%%%%%%%%%%%%%%%%%%%%%%%%%%%%%%%%%%%%%%%

%%%%%%%%%%%%%%%%% BODY OF PAPER %%%%%%%%%%%%%%%%%%
\section{Introduction}

Within the next few years, the eROSITA X-ray telescope on board the Spectrum-Roentgen-Gamma (SRG) mission will detect and measure the position and properties of about 3 million active galactic nuclei (AGN) and 100\,000 clusters of galaxies over the full sky \citep{arxiv12093114_Merloni,2016SPIE.9905E..1KP}. 
Such large numbers of AGN will enable us to compute accurately at least their 1- and 2-point distribution functions over cosmological volumes, and thereby tightly interconnect the evolution of growing supermassive black holes in the overall galaxy population with the large scale structure of dark matter \citep{Fanidakis2011MNRAS41053F}. 
For example, the upcoming clustering measurements of the most luminous AGN over the full sky should be precise enough to retrieve their corresponding halo population parameters to within better than 30\% ($\sim$0.1 dex on halo masses).

To enable these clustering measurements, large-scale cosmological simulations are a key tool for three main reasons: 
Firstly, they enable a thorough validation of the method used to construct the catalogues and understand possible systematic errors. 
They constitute an input for end-to-end simulations of the catalogue creation process \citep[\textit{e.g.}][]{Reid2016MNRAS4551553R}. 
Secondly, they allow the construction and constraint of the possible halo occupation distribution of AGN and represent a key tool to assess possible systematic errors on the measurements \citep[\textit{e.g.}][]{ross2017MNRAS4641168R}.
Finally, they provide the necessary information to create covariance matrices with respect to cosmological parameters or halo occupation distribution parameters \citep[\textit{e.g.}][]{Klypin2018MNRAS4784602K}. 

This paper is a continuation of \citet[][hereafter G18]{2018MNRAS.tmp.3272G}, where a model linking AGN to dark matter haloes was explained and tested against a set of clustering measurements of X-ray selected AGN available in the literature. 
We present here a generalization of this method to redshift $0<z<6$ that we project on the full sky by including an accurate model of the eROSITA all-sky survey selection function. 
The mock catalogue presented here is a cornerstone to start the eROSITA end-to-end data processing validation tasks. 

The structure of the paper is as follows: We discuss the cosmological N-body simulations used to create the mock in Sec.~\ref{sec:nbody:data}. 
Then, we describe the analytic model we created to simulate the AGN population in Sec.~\ref{sec:agn:model}; 
we explain how AGN populate dark matter haloes and how their X-ray emission is calculated; we detail the dust extinction model used to predict both the X-ray and the optical emission of AGN. 
In Section \ref{sec:results}, we present a set of forecasts generated using the mock catalogue itself. 

First we give the expected redshift distribution of eROSITA AGN as a function of survey depth and AGN type.
Then, we lay out a possible spectroscopic follow-up strategy (see Sec. \ref{subsection:specz:follow:up}). 
In Sec.~\ref{subsection:XLF:prediction}, we detail how well the X-ray luminosity function will be recovered at the bright end, while in Sec. \ref{subsection:clustering} we forecast the accuracy with which the baryon acoustic oscillation standard ruler will be measured throughout the redshift range $0.5<z<4$. In \ref{subsec:ssc}, we detail to what accuracy the large scale halo bias of AGN will be measured. 
In Sec. \ref{sec:conclusion}, we detail a possible strategy towards completing eROSITA end-to-end simulations. 

We assume a flat $\Lambda CDM$ cosmology close to that of the \citet{Planck2014} with $\Omega_m=0.307115$,  $h=0.6777$, $\sigma_8=0.8228$. Magnitudes are in the AB system. 
Throughout the paper $L_X$ refers to the X-ray luminosity in the 2-10 keV band.
The set of eROSITA mocks is made public here\footnote{\url{http://www.mpe.mpg.de/~comparat/eROSITA_AGN_mock/}}

\section{N-body data}
\label{sec:nbody:data}

\begin{figure*}
\centering
\includegraphics[width=16cm]{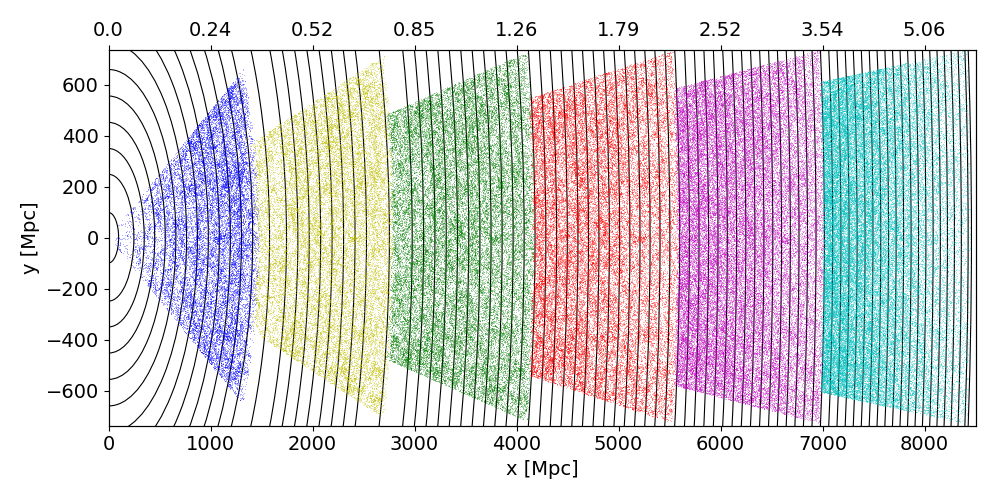}
\caption{Layout of the box replication scheme.  
The six line of sight replications are shown in different colors, they pave up to 8500 comoving Mpc away from us, which corresponds to the redshift interval 0 to 6 (top ticks). 
Each dot represents a dark matter halo (only a small fraction of the haloes are represented, the thickness along the third axis is about 20Mpc). 
The black lines correspond to boundaries between the shells i.e. the discrete time steps of evolution of halos.}
\label{fig:box:replication}
\end{figure*}

\begin{table*}
	\centering
	\caption{Light cone shells specifications. 
	Area: sky coverage subtended (before replication). 
	R.A. and Dec. : angular size (or opening angle). 
	d (min,max): minimum and maximum comoving distance to the observer. 
	z (min,max): minimum and maximum redshift to the observer. 
	N snapshots: number of snapshots used. 
	N distinct: Number of distinct haloes contained in the projected simulation within the R.A., Dec. and redshift boundaries that are more massive than $M_{vir}\geq 2\times 10^{11} M_\odot$ ($\sim$100 bound particles per halo). 
	N replications: Number of replications needed to pave the full sky.
	}
	\label{tab:light:cones}
	\begin{tabular}{lr cc rrr rrr} 
	\hline
name  & area & R.A. & Dec. & d (min,max) &  z (min,max) & N snapshots & N distinct & N replications\\%& l o s stretch \\%& N haloes \\
          & [deg$^2$] & [deg] & [deg] & [Gpc] & & & halos\\
          \hline
C1    & 3044 & 60.0 & 53.0 & (0.000, 1.475) & (0.000, 0.365) & 16 & 6836290  & 21 \\ 
C2    & 894  & 30.6 & 29.4 & (1.475, 2.795) & (0.365, 0.773) & 13 & 11690505 & 91 \\ %      
C3    & 405  & 20.3 & 20.0 & (2.795, 4.178) & (0.773, 1.344) & 15 & 13938405 & 171 \\ %
C4    & 227  & 15.1 & 15.0 & (4.178, 5.590) & (1.344, 2.192) & 16 & 13117165 & 325 \\ %
C5    & 144  & 12.1 & 12.0 & (5.590, 7.017) & (2.192, 3.559) & 18 & 8520481  & 465 \\ %
C6    & 99   & 10.0 & 9.8 & (7.017, 8.454) & (3.559, 6.000) & 21 & 2237965  & 703 \\ %
\hline
	\end{tabular}
\end{table*}

We use the halo catalogues of the MDPL2 simulation\footnote{\url{https://www.cosmosim.org/cms/simulations/mdpl2}} (Planck cosmology, $3840^3$ particles of mass $2.2\times10^9 M_\odot$, 1475.5 Mpc on the side, \citealt{Klypin2016}) post-processed with the \textsc{rockstar} merger trees software \citep{Behroozi2013}. 
The halo mass function is correct to a few percent down to halo masses of $M_{vir}\sim10^{11}M_\odot$ \citep{2017ComparatHMF}. 
In this analysis, we consider distinct haloes more massive than $2\times 10^{11} M_\odot$. 
This resolution is sufficient to resolve haloes that contain the least massive AGN to be observed by eROSITA; indeed, G18 showed that a higher resolution simulation would not bring additional information on cosmological scales. 
Alternative simulations which could be used for this project include those of \citet[\textit{e.g.}][]{Angulo2012,Skillman2014,Heitmann2015,Ishiyama2015}.

The configurations used to cover the redshift range $0<z<6$ are given in Table \ref{tab:light:cones}. 
We replicate 6 times the box along the line of sight to reach up to redshift 6. 
Each replicated box is sliced (like an onion) into shells according to the availability of snapshots. 
There are about 20 snapshots per replica, implying an approximate width of about 80 Mpc per snapshot. 
This corresponds to a time step between shells of 50Myr at redshift 6 that increases to 250Myr at redshift 0.
The shells and replication scheme are illustrated in Fig. \ref{fig:box:replication}.
We then pave these shells on the full sky using rotation matrices. 
We compute the new position and velocity vectors for each halo and its projection in right ascension, declination and redshift (real and redshift space). We then populate or `paint' the full sky mock with AGN, as we describe in the following section. 

\section{Active galactic nuclei model}
\label{sec:agn:model}
We create an analytic model of the population of X-ray emitting supermassive black holes, which we subsequently refer to as `AGN' (see \citet{padovani2017AAR} for extended discussions about the AGN naming conventions). 
The aim of this model is to follow accurately, within 10-20\%, the observed hard X-ray luminosity function and the number density projected on the sky as a function of soft X-ray flux, while tracing the large-scale structure of the universe. 

We use as starting point semi-analytical models (SAMS, \textit{e.g.} \citealt{2016MNRAS.462.3854L}), which reproduce quite accurately the bulk of the galaxy population. 
AGN, however, are not typically accurately characterized in SAMS. 
It is indeed still a matter of debate how supermassive black holes evolve within the galaxy population, and physical prescriptions struggle to predict jointly within 10\% the galaxy and AGN population \citep{Caplar2015ApJ...811..148C,Caplar2018ApJ...867..148C,Mayer2019RPPh82a6901M}. 

Hydro-dynamical simulations, coupled with sub-grid prescriptions, are able to reproduce the X-ray luminosity function of AGN \citep[\textit{e.g.}][]{2017MNRAS.468.3395M,biffi_2018MNRAS.tmp.2317B}. 
However, due to computational limitations, they span volumes that contain only small numbers of AGN and thus cannot provide the accuracy needed over the volume considered here (full sky), in particular, at the bright end of the luminosity function \textit{e.g.} \citet{2014MNRAS.442.2304H}, Fig. 9 or \citet{Koulouridis2018AA620A4K}, Fig. 4.

The model created here is empirical, and follows the principles outlined in \citet{Croton2009MNRAS.394.1109C}. 
It is constructed to reproduce accurately the observed number densities of AGN as a function of observed X-ray luminosity and redshift. 
Methodologically, it resembles the halo abundance matching technique used to create high fidelity galaxy mock catalogs, which has been extensively adopted in recent years \citep{2004ApJ...609...35K,2006ApJ...647..201C,2010MNRAS.404.1111G,2010ApJ...717..379B,2011ApJ...742...16T,2015MNRAS.447.3693K,2016MNRAS.461.3421F,2016MNRAS.460.1173R,sergio_2017MNRAS468728R,2018arXiv181005318G}. 

Our model goes beyond (or differs from) G18 in the following aspects:
\begin{itemize}
 \item It is a generalization to all snapshots with redshift $z<6$ re-processed into a full-sky light cone (instead of 3 snapshots sampling discrete redshift values 0.2, 0.75 and 1). 
\item We moved away from sampling specific accretion rates and stellar masses from observed distributions, as in \citep[][G18]{Bo2016AA588A78B,2018MNRAS.tmp.3272G}. 
In the full sky simulation, the many haloes made this computationally too slow. 
Instead, we created a (much faster) abundance matching scheme that reproduces the luminosity and mass functions as well as the specific accretion rate distributions of super massive black holes.
 \item Our model follows the hard X-ray luminosity functions from \citet{2015MNRAS.451.1892A} to better than $\pm5\%$. 
 \item An eROSITA sensitivity model for its all sky survey components is included by following the descriptions from \citet{2018AA...617A..92C}. 
 \item Multi-wavelength properties and classification of AGN are computed: broad-band magnitudes (ultra-violet, optical and infrared), plus medium resolution optical spectra.
\end{itemize}

\subsection{Abundance matching model}

\subsubsection{Galaxy stellar masses}

\begin{figure}
\centering
\includegraphics[width=.82\columnwidth,type=png,ext=.png,read=.png]{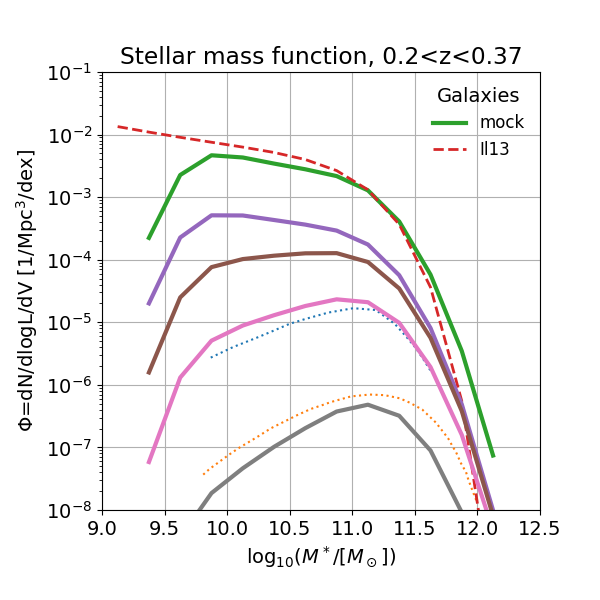}
\includegraphics[width=.82\columnwidth,type=png,ext=.png,read=.png]{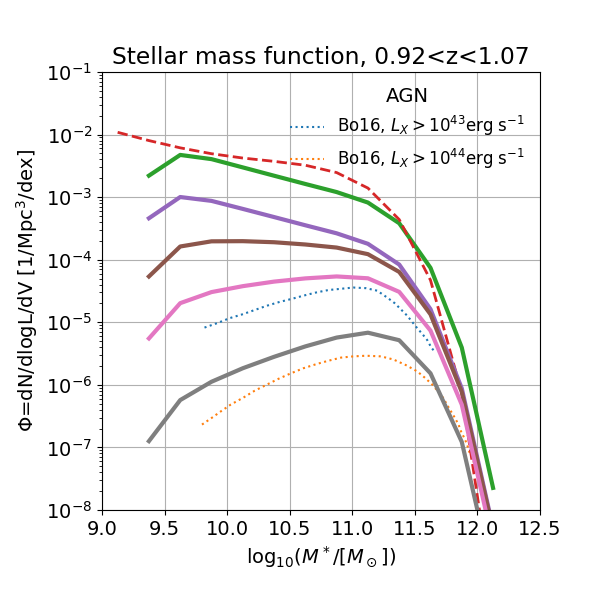}
\includegraphics[width=.82\columnwidth,type=png,ext=.png,read=.png]{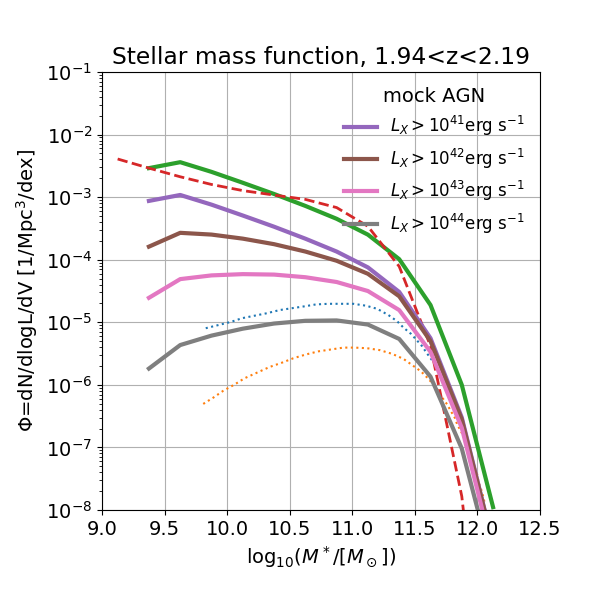}
\caption{Stellar mass function (SMF) at redshift 0.3, 1, and 2. 
The SMF sampled by the mock galaxies is shown with a green solid line. 
It is in qualitative agreement with the SMF measured in the COSMOS survey \citep[red dashed line,][]{Ilbert2013AA556A55I}. 
The SMF of mock galaxies hosting AGNs (solid lines) are shown for 4 threshold of hard X-ray luminosity 	: greater than $10^{41}$, $10^{42}$, $10^{43}$, $10^{44}$ erg s$^{-1}$. 
The observed SMF of galaxies hosting AGNs (dotted lines) from \citet{Bo2016AA588A78B} are within a factor of 2-5 from the mock.
}
\label{fig:SMF}
\end{figure}

We use the stellar mass to halo mass relation from \citet{2013MNRAS.428.3121M} to assign all distinct haloes a stellar mass $M^*$ for their central galaxy. 
The logarithm of the stellar masses are drawn from a Gaussian distribution (with scatter 0.15 dex) around a mean obtained with Eq. \ref{eqn:SMHMR}. 
\begin{equation}
\label{eqn:SMHMR}
 \bar{M}^*(M_h, z) = \frac{A(M_H,z)}{B(M_h,z) + C(M_h,z)} 
\end{equation}
where $M_h=M_{vir}$ and 
\begin{align*}
& A(M_h,z) = 4 M_h \left(0.0351 - 0.0247 \frac{z}{1+z}\right) \\
& B(M_h,z) = \left(\frac{M_h}{10^{11.59 + 1.195  \frac{z}{1+z} } } \right)^{- 1.376 + 0.826  \frac{z}{1+z}} \\
& C(M_h,z) = \left( \frac{M_h }{10^{11.59 + 1.195 \frac{z}{1+z}} }\right)^{0.608 + 0.329 \frac{z}{1+z}}.
\end{align*}
We then select haloes with $M_{vir}>2\times10^{11} M_\odot$ that host galaxies with a mass $M^*>2\times10^9 M_\odot$. 
The stellar mass function obtained is in good qualitative agreement with the literature \citet[\textit{e.g.}][]{Ilbert2013AA556A55I,2017AA...605A..70D}. The obtained stellar mass function starts to turn over (i.e become incomplete) at around $10^{10} M_\odot$ at low redshift ($z<1$) and $10^{9.5} M_\odot$ at high redshift. 
In Fig. \ref{fig:SMF}, we show how the stellar mass function of all mock galaxies compares with the observed stellar mass function at a similar redshift measured in the COSMOS field. 
Overall the mock stellar mass function is within a factor of 3 of \citet{Ilbert2013AA556A55I}. At the high mass end, it can differ by a factor of 20.
We refrain from fine-tuning the stellar-to-halo mass relation to obtain the closest possible agreement with data.
The SMF of galaxies hosting AGNs present in the mock is within a factor 2-5 of observations. 

\subsubsection{AGN Duty cycle}
\begin{figure}
\centering
\includegraphics[width=.81\columnwidth,type=png,ext=.png,read=.png]{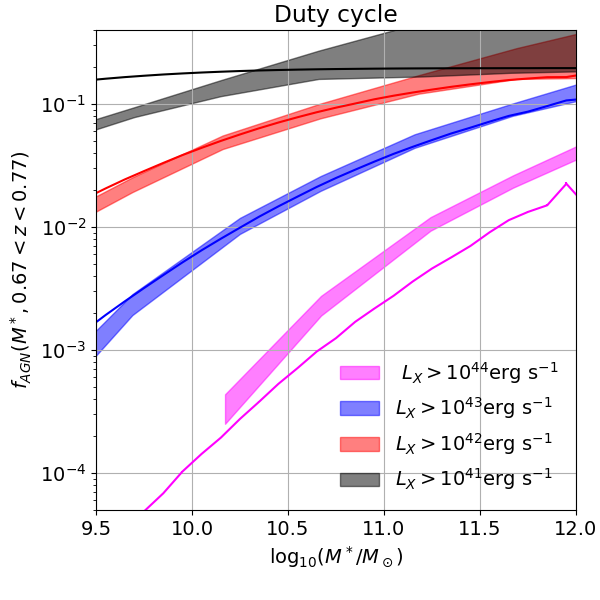}
\includegraphics[width=.81\columnwidth,type=png,ext=.png,read=.png]{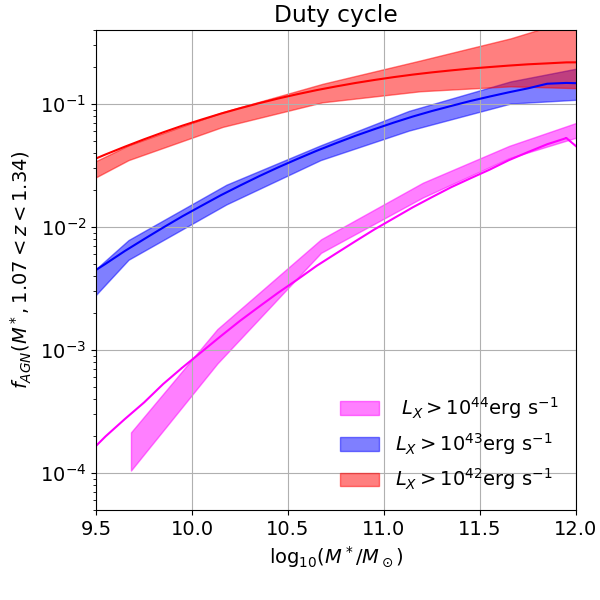}
\includegraphics[width=.81\columnwidth,type=png,ext=.png,read=.png]{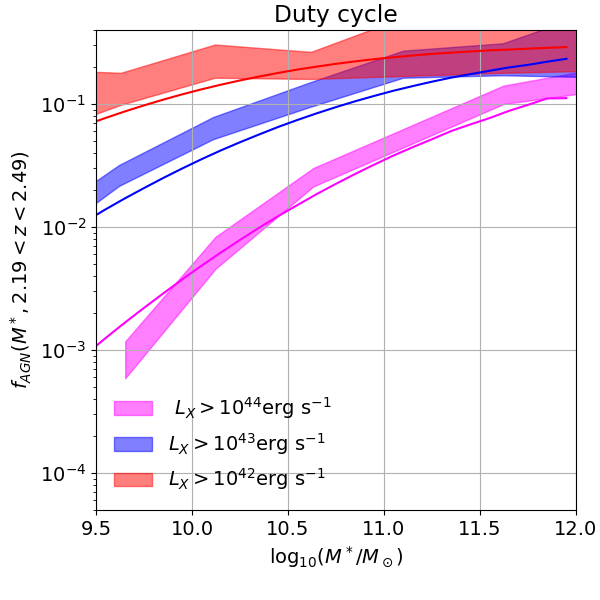}
\caption{Fraction of active galaxies (Duty cycle, DC) brighter than a given hard X-ray luminosity (2-10 keV band) threshold as a function of stellar mass at redshift 0.7, 1,2, 2.3. 
The DC sampled by the mock (solid lines) is in good qualitative agreement with the measurements of \citet{2017MNRAS.471.1976G} at redshifts 0.75, 1.25, 2.25 (shaded areas). 
The maximum values at the high mass end are in agreement by construction because they are input in the model. 
How the DC decreases with stellar mass for a luminosity threshold is related to the scatter in the abundance matching relation. 
}
\label{fig:duty:cycle}
\end{figure}

We use the AGN duty cycle values (i.e. the fraction of active galaxies at any given stellar mass and redshift) measured in \citet{2017MNRAS.471.1976G}, taking these from the top-right corner of each panel of their Fig. 14.  
We linearly interpolate between the duty cycle values (0.1, 0.2, 0.3, 0.3) at redshifts (0., 0.75, 2., 6.1), respectively. This means that, \textit{e.g.}, at redshift 0 (2), 10\% (30\%) of the haloes are randomly chosen and designated as AGN hosts. 
Note that \citet{2017MNRAS.471.1976G} only samples the duty cycle out to z=4, so we extrapolate the z=4 duty cycle value up to z=6. 
Our model of the duty cycle is shown in Fig. \ref{fig:duty:cycle} as a function of stellar mass, and it agrees well with the measurements of \citet{2017MNRAS.471.1976G}. 
The dependence of the duty cycle on the stellar mass for a given X-ray luminosity threshold is a consequence of the scatter in the abundance matching relation between stellar mass and X-ray luminosity (see Sec. \ref{subsubsec:HAM}). 
Fig. \ref{fig:SMF} shows how the duty cycle projects on the stellar mass function. 
The curves corresponding to luminosity thresholds of $10^{43}$ and $10^{44}$ based on the mock are within a factor of 2-5 of the model of \citep{Bo2016AA588A78B}. 
Note that we do not assume a functional form for the duty cycle as a function of stellar mass, neither do we seed black holes in the halos. 

\subsubsection{Hard X-ray luminosity distributions}
\label{subsubsec:HAM}
\begin{figure}
\includegraphics[width=.83\columnwidth,type=png,ext=.png,read=.png]{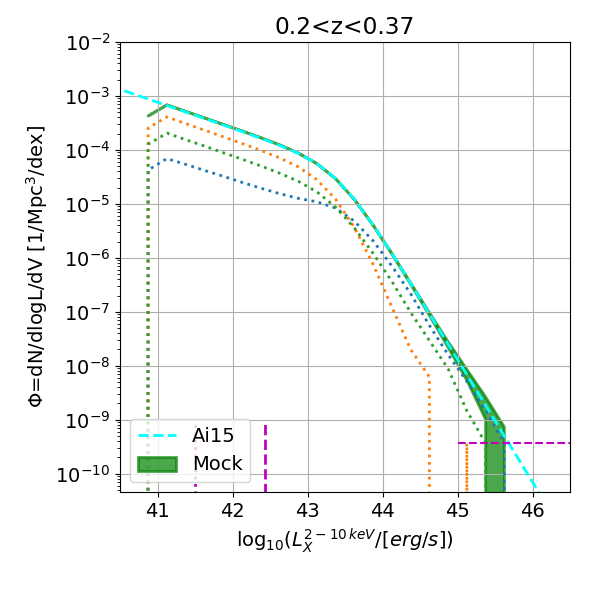}
\includegraphics[width=.83\columnwidth,type=png,ext=.png,read=.png]{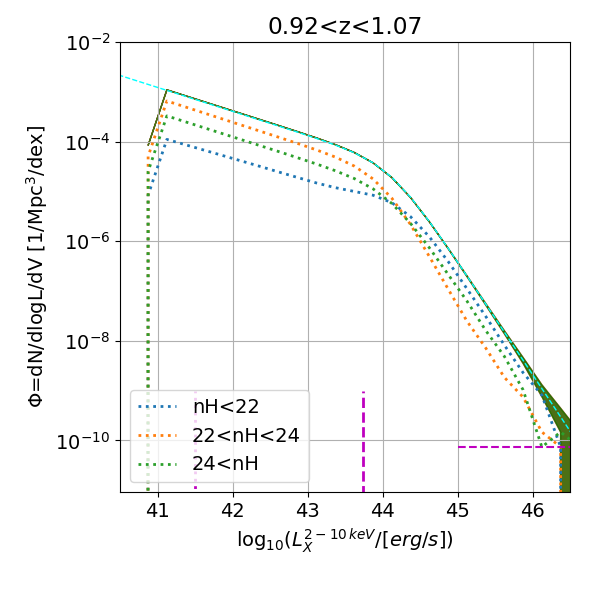}
\includegraphics[width=.83\columnwidth,type=png,ext=.png,read=.png]{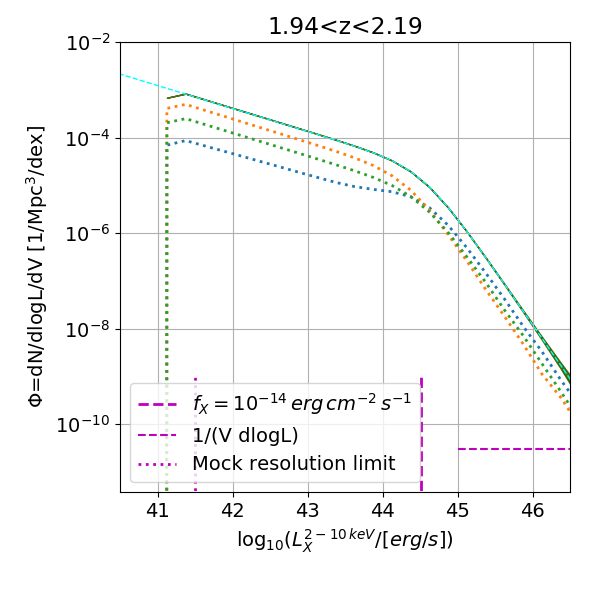}
\caption{Hard XLF of the mock: number density v.s. hard X-ray luminosity (2-10 keV band) for three redshift bins: 0.25, 1., 2 (green contour). 
This can be compared to the model from \citet{2015MNRAS.451.1892A} (blue dashed line). 
The decomposition of the XLF into three obscuration bins ($n_H$) is shown as the dotted lines. 
The inverse of the volume of the redshift shell is shown by the line dubbed '1/(V dlogL)' (horizontal magenta thin dashes). 
Below this limit Poisson noise dominates. 
The mock resolution limit is around $L_X=10^{41.5}$erg s$^{-1}$ (vertical magenta dots). 
The vertical magenta thick dashes give the luminosity that if emitted at the mean redshift of the bin, would result in an observed hard X-ray flux of $\sim10^{-14}$ erg cm$^{-2}$ s$^{-1}$.
}
\label{fig:XLF}
\end{figure}

We adopt the AGN hard X-ray luminosity function (XLF) from \citet{2015MNRAS.451.1892A} and specifically the LADE model therein. This provides a good (and simple) description of the XLF up to redshift 5. 
For our needs, we extrapolate the 2-10 keV XLF models up to redshift 6. 
From the hard XLF model, we draw the set of luminosities, $L_X$, to be attached to the AGNs in redshift shells over the full sky. 
We create an intermediate variable 
\begin{equation}
\tilde{M^*} = \log_{10}(L_X/[erg\; s^{-1}]) + N(0,\sigma) 
\end{equation} 
that contains a Gaussian scatter that varies from 1.4 dex at redshift 0 to 1 dex at redshift 6 as follows:
\begin{equation}
\label{eqn:scatter:HAM}
\sigma(z)=1.4 -2z/30. 
\end{equation}
We rank order match the stellar masses to the intermediate variable: the largest $M^*$ is connected to the largest $\tilde{M^*}$ and assigned the corresponding $L_X$. 
Adding scatter in this fashion is computationally very efficient compared with sampling from the observed bivariate distributions of stellar mass and specific accretion rate \citep{Bo2016AA588A78B,2018MNRAS.tmp.3272G}. 
This procedure guarantees the reproduction of the hard X-ray (2-10keV) luminosity function (see Fig. \ref{fig:XLF}). 
Indeed, the mock catalogue's XLF is within $\pm5\%$ of the model. 
The AGN host galaxy stellar mass function for hard X-ray luminosity thresholds is shown in Fig. \ref{fig:SMF}: it is in fair agreement with the model of \citet{Bo2016AA588A78B}. 

\begin{figure}
\includegraphics[width=.83\columnwidth,type=png,ext=.png,read=.png]{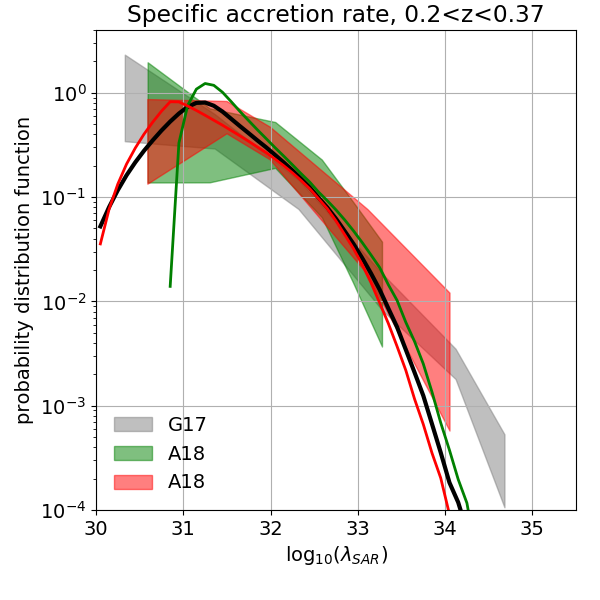}
\includegraphics[width=.83\columnwidth,type=png,ext=.png,read=.png]{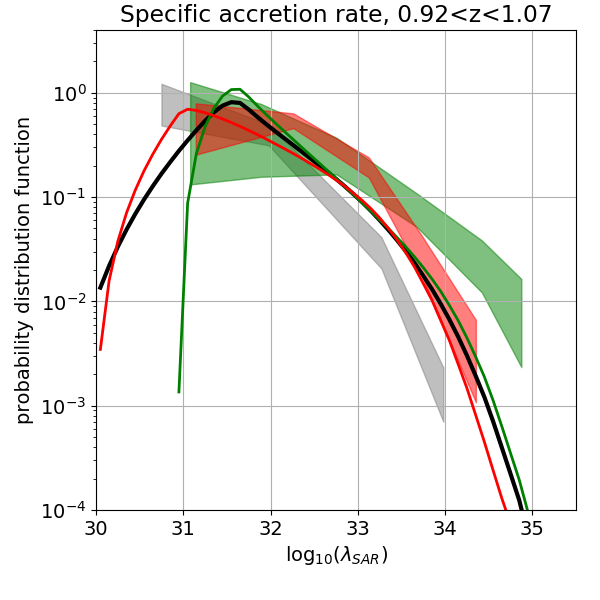}
\includegraphics[width=.83\columnwidth,type=png,ext=.png,read=.png]{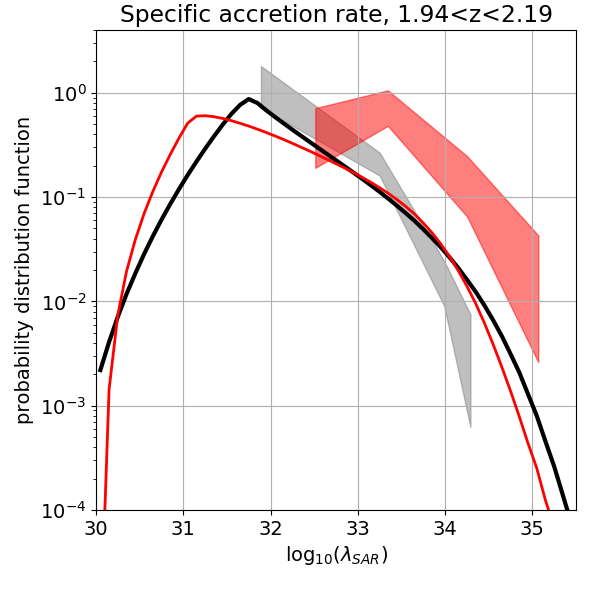}
\caption{Specific accretion rate ($\lambda_{SAR}$) probability distribution function of the mock (solid lines) and from observations (shaded areas).  
The complete set  ($8<\log_{10}(M_*/M_\odot)$) is shown in black/gray. 
Subsets in bins of stellar mass $9<\log_{10}(M_*/M_\odot)<10$ ($10<\log_{10}(M_*/M_\odot)<11$) are in green (red). 
It is key to reproduce the slope i.e. to sample a broad range of $\lambda_{SAR}$.
The slope obtained in the mock compares well to the ones measured in \citet{2017MNRAS.471.1976G} (G17) \citet{2018MNRAS.474.1225A} (A18), although the observed distributions are slightly broader than those in the mock. 
}
\label{fig:LSAR}
\end{figure}

\subsubsection{Specific accretion rate}
The ratio between hard X-ray luminosity and stellar mass gives the specific accretion rate, hereafter $\lambda_{SAR}$[erg s$^{-1}M_\odot^{-1}$]$=L_X/M^*$. 
We compare the distributions we obtain to current estimates from \citep{2017MNRAS.471.1976G,2018MNRAS.474.1225A} on Fig. \ref{fig:LSAR}. 
The $\lambda_{SAR}$ distribution is directly related to the scatter in Eq. \ref{eqn:scatter:HAM}. 
A large scatter, \textit{e.g.} 1.5, produces a $\lambda_{SAR}$ distribution that will be broad (flat), as observed in the data. 
A small scatter, \textit{e.g.} 0.2, produces a very narrow $\lambda_{SAR}$ rate distribution (i.e. a 'light-bulb' model for AGN activity, unlike in the data). 
A scatter parameter between 1 and 1.5 produces reasonable distributions $\lambda_{SAR}$ and host stellar mass. 
It is key to reproduce the slope i.e. to sample a broad range of $\lambda_{SAR}$.
The mock compares well to \citet{2017MNRAS.471.1976G} and \citet{2018MNRAS.474.1225A} at low redshift at all specific accretion rates, although the observed distributions are slightly broader than in the mock at the high $\lambda_{SAR}$ end. 
At high redshift, the slope in the mock is still in good agreement while the normalization is off. 
Note that, because of the large uncertainties on the measurements and the differences in coverage of $\lambda_{SAR}$ between the mock and the data, the normalization of the probability distribution function (done on the mean value, only where the data are not consistent with 0) can be off by a factor of a few in the y-axis. 
To obtain a similar normalization, we would need to model in detail the selection functions of \citet{2017MNRAS.471.1976G} and \citet{2018MNRAS.474.1225A}, which is beyond the scope of this paper. 

Increasing the resolution of the simulation, i.e. including lower mass galaxies further broadens the $\lambda_{SAR}$ at the high $\lambda_{SAR}$ end. For a galaxy that is assigned the same luminosity, its stellar mass may be smaller and its specific accretion rate higher. 

\subsubsection{Limitation}

The main limitation of the abundance matching models come from the limited resolution of the simulation: the faint luminosity limit depends on redshift. At low redshift $z<0.1$, we are missing some low-luminosity AGN. 

\subsection{Obscuration model}
\label{sec:obscuration}
We create an obscuration model for the AGN similar to the ones proposed by \citet{2017Natur.549..488R, 2017MNRAS.465.4348B}. 
We add a fine redshift variation to it, following observations from \citet{2014ApJ...786..104U, 2015MNRAS.451.1892A, 2015ApJ...802...89B}. 
We consider three bins in the logarithm of the absorbing (neutral) column density ($n_H$): 20-22 unobscured, 22-24 Compton-thin obscured (CTN), 24-26 Compton-thick obscured (CTK). 
This model gives, as a function of hard X-ray luminosity and redshift, the fraction of AGN in each obscuration category. 

The fraction of CTK AGN is fixed at 30\%.

\begin{equation}
 f_{\rm thick} = 0.3 
\end{equation}
The fraction of CTN plus CTK AGN depends on redshift and hard X-ray luminosity 
\begin{align}
\label{eqn:obscuration}
 f_{\rm thin+thick}(l_x, z) = 
 & f_1(z)+(f_2(l_x)-f_1(z) )\times \cdots \\ 
 &\cdots \left[\frac{1}{2}+\frac{1}{2}\epsilon{\left( \frac{LL(z)-l_x}{0.6} \right)}\right] 
\end{align}
where 
\begin{align}
\label{eq:def:lx:logLX}
& l_x = \log_{10}(L_X/{\rm [erg\; s^{-1}]}), \\
& f_1(z) = f_{\rm thick} + 0.01 + 0.4\epsilon(z/4), \\
& f_2(l_x) = 0.9\sqrt{41/l_x}, \\
& LL(x) = 43.2 + 1.2 \epsilon(x)
\end{align}
and $\epsilon$ is the Gauss error function 
\begin{equation}
\label{eqn:gauss:err:fct}
\epsilon(x)=\frac{2}{\sqrt{{\rm \pi}}}\int_{t=0}^{t=x} {\rm e}^{-t^2}\,dt
\end{equation}

\begin{figure}
\begin{center}
\includegraphics[width=0.8\columnwidth,type=png,ext=.png,read=.png]{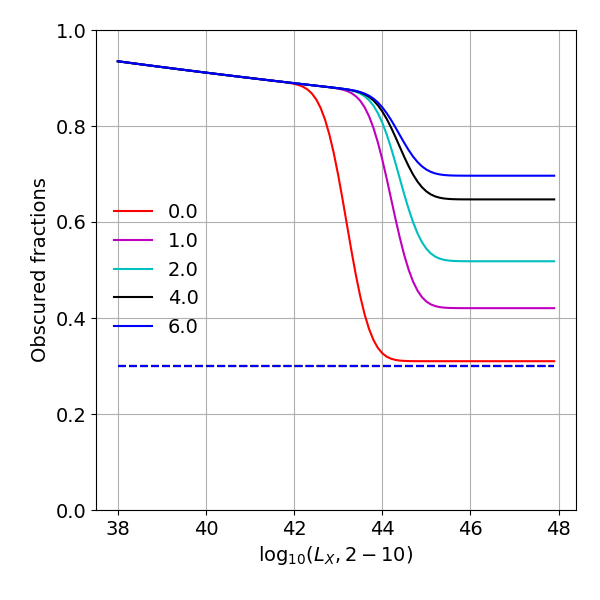}
\end{center}
\caption{Model of the obscured fraction of AGN as a function of hard X-ray luminosity (2-10 keV band) at different redshifts. The blue dashed line shows the fraction of CTK objects: 0.3. The solid lines show the summed fractions of CTK plus CTN AGN at redshifts 0, 1, 2, 4, 6. 
}
\label{fig:model:obscuration}
\end{figure}

Fig. \ref{fig:model:obscuration} shows how Eq. \ref{eqn:obscuration} evolves as a function of $L_X$ and redshift. It is similar to that of \citet{2017Natur.549..488R, 2017MNRAS.465.4348B}.

\begin{figure}
\begin{center}
\includegraphics[width=0.86\columnwidth,type=png,ext=.png,read=.png]{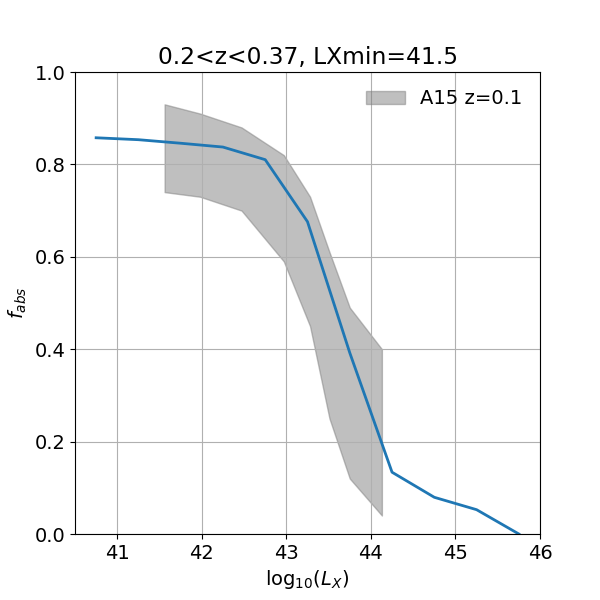}
\includegraphics[width=0.86\columnwidth,type=png,ext=.png,read=.png]{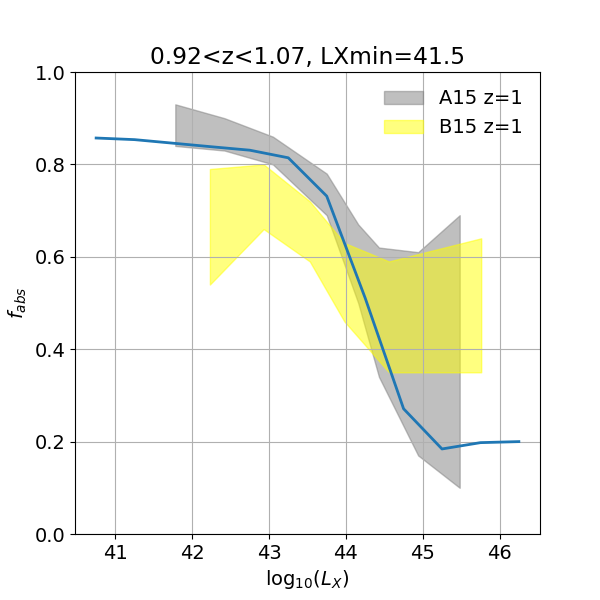}
\includegraphics[width=0.86\columnwidth,type=png,ext=.png,read=.png]{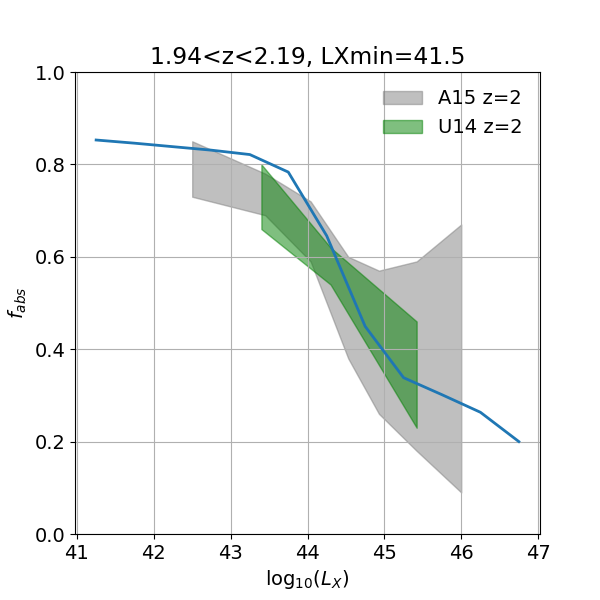}
\end{center}
\caption{$f_{abs}$ as a function of hard X-ray luminosity(2-10 keV band) at three redshifts: 0.25, 1, 2. 
It is compared to its equivalent in the observations  \citep[U14][]{2014ApJ...786..104U}, \citep[A15][]{2015MNRAS.451.1892A}, \citep[B15][]{2015ApJ...802...89B}.
}
\label{fig:model:f_abs:LX}
\end{figure}

\begin{figure}
\begin{center}
\includegraphics[width=0.85\columnwidth,type=png,ext=.png,read=.png]{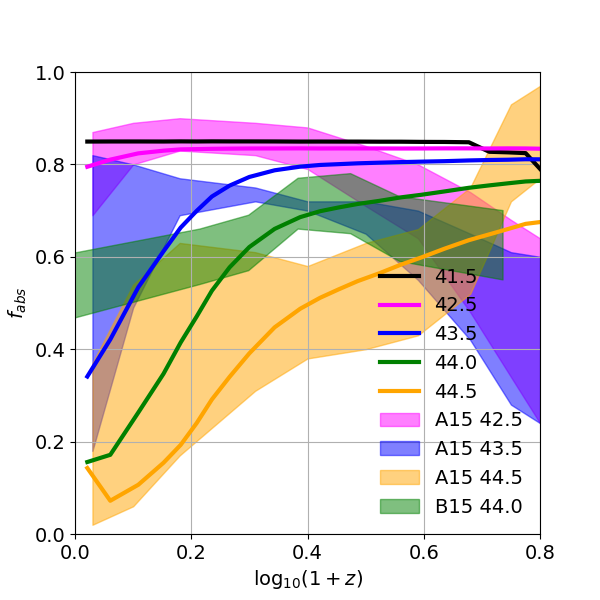}
\includegraphics[width=0.85\columnwidth,type=png,ext=.png,read=.png]{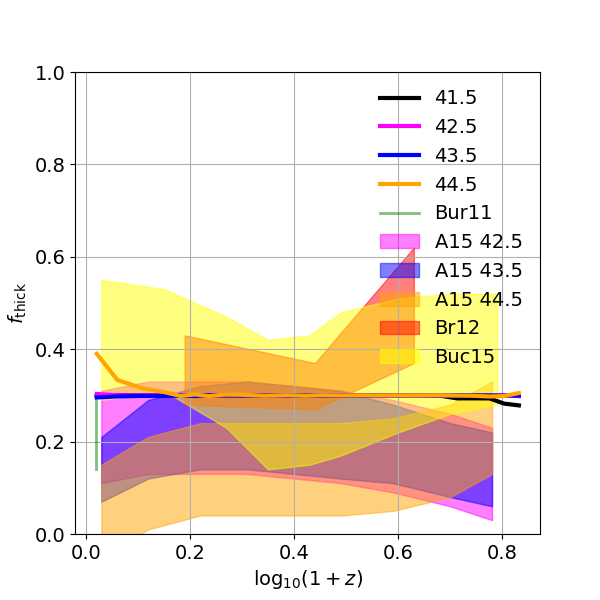}
\end{center}
\caption{$f_{abs}$ ($f_{\rm thick}$) as a function of redshift for a set of hard X-ray luminosity (2-10 keV band) thresholds, top (bottom) panel.}
\label{fig:model:f:abs:f:tk:z}
\end{figure}

We run one random number per AGN, denoted $\mathcal{R}$. 
It is uniformly distributed between 0 and 1. 
We use it to assign the $n_H/[\mathrm{cm}^{-2}]$ values to each AGN. 
For each AGN we evaluate $f_{\rm thin+thick}(l_x,z)$ using Eq. (\ref{eqn:obscuration}) and compare it to the random number. 
If $\mathcal{R} \geq f_{\rm thin+thick}(l_x,z)$, the AGN is classified as unobscured and we assign $\log_{10}(n_H)$ randomly in the interval 20 (included) and 22 (excluded). 
If $\mathcal{R} \leq f_{\rm thick}=0.3$, the AGN is classified as thickly obscured and we assign $\log_{10}(n_H)$ randomly in the interval 24 (included) and 26 (excluded). 
Else, the AGN is thinly obscured, we assign $\log_{10}(n_H)$ randomly in the interval 22 (included) and 24 (excluded).
\begin{align}
 \log_{10}(n_H) \sim Uniform(20,22) & \; {\rm if}\; \mathcal{R}  \geq f_{\rm thin+thick} \\
 \log_{10}(n_H) \sim Uniform(24,26) & \; {\rm if}\; \mathcal{R}  \leq f_{\rm thick}      \\
 \log_{10}(n_H) \sim Uniform(22,24) & \; {\rm otherwise} 
\end{align}
The fraction $f_{abs}$ defined as 
\begin{equation}
f_{abs} = \frac{N(22<\log_{10}(n_H)<24)}{N(20<\log_{10}(n_H)<24)}
\end{equation}
thus depends on luminosity. 
The decomposition of the soft XLF into the different populations of unobscured, obscured and thick obscured AGN is shown on Fig. \ref{fig:XLF}.
We show the obtained $f_{abs}$ as a function of hard X-ray luminosity in Fig. \ref{fig:model:f_abs:LX}, it is compared to observational results used to design the model \citep{2014ApJ...786..104U, 2015MNRAS.451.1892A, 2015ApJ...802...89B}. 
Fig. \ref{fig:model:f:abs:f:tk:z} shows the evolution of $f_{abs}$ and $f_{\rm thick}$ with redshift for a set of luminosity thresholds. 
The measurement of the obscured and CTK fraction is model dependent, so, to be consistent with the XLF from \citet{2015MNRAS.451.1892A} used in the abundance matching procedure, our obscured fraction model follows their constrains up to redshift $\sim2$, see Fig \ref{fig:model:f_abs:LX}. 
At higher redshifts ($z>2$ or $\log_{10}(1+z)>0.5$) the data do not constrain well the obscured fractions, \citet{2014ApJ...786..104U,2015MNRAS.451.1892A, 2015ApJ...802...89B} find CTK fractions and obscured fractions that are in tension with each other. 
For the obscured fraction, for simplicity, we choose not to add a component to the model to reproduce a tentative decrease in the obscured fraction at high redshift. 
So it may well be that our model produces too many obscured AGNs at high redshift. 
eROSITA data will constitute a great benchmark for the bright and high redshift end of the obscuration model. 
Based on observations shown on Fig. \ref{fig:model:f:abs:f:tk:z}, the fraction of CTK AGNs could vary between 10 and 50\%. 
For simplicity, we use a constant fraction set at 30\%. 

\subsection{Soft and hard X-ray flux}

\begin{figure}
\begin{center}
\includegraphics[width=0.8\columnwidth,type=png,ext=.png,read=.png]{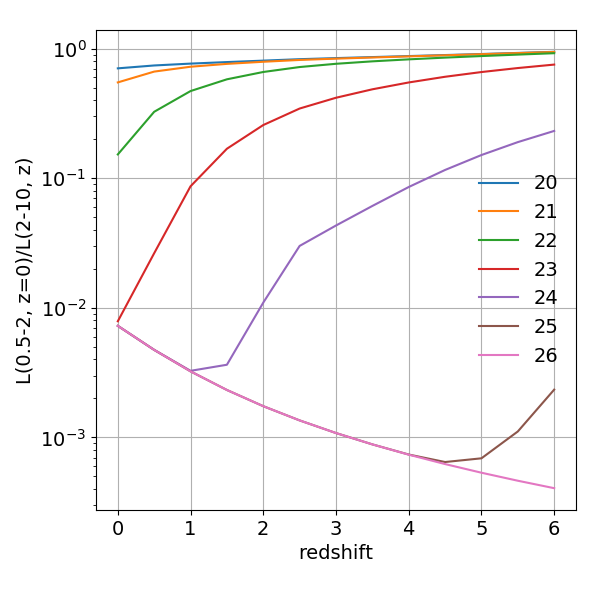}
\includegraphics[width=0.8\columnwidth,type=png,ext=.png,read=.png]{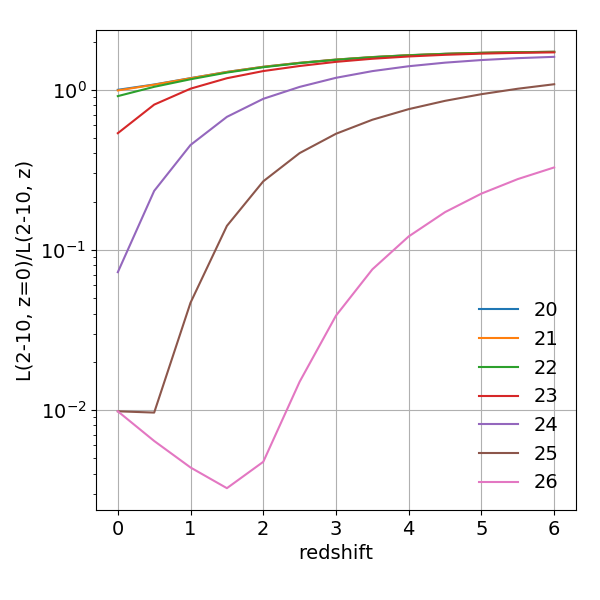}
\end{center}
\caption{Conversion from the rest-frame hard X-ray luminosity (2-10 keV band) to fluxes in the soft (top panel) and hard (bottom panel) bands as a function of redshift for a discrete set of $\log_{10}(n_H/\mathrm{cm}^{-2})$ values.}
\label{fig:model:kcorrection}
\end{figure}

To convert the hard X-ray luminosities to hard and soft X-ray fluxes, we use the obscuration model combined with a set of X-ray spectra.
With the $n_H$ values, we convert the rest-frame $L_{X,\; 2-10keV}$ to an observed-frame flux in the soft and hard X-ray bands: 0.5-2keV and 2-10keV. 
We tested two models of AGN spectra, which provide almost indistinguishable results. 
The first one is a Clumpy torus model with reflection, see \citet{2018cosp...42E.456B}. 
The second is the one used in \citet{2015MNRAS.451.1892A} i.e. an absorbed power-law with Compton reflection and a soft scattered component at the 2\% level. 
For consistency with the XLF (also taken from \citet{2015MNRAS.451.1892A}), we implemented the second set of spectra in the mock catalog. 
Fig. \ref{fig:model:kcorrection} shows the combined effects of the K-correction (redshift effect) and band conversion (from hard 2-10 keV to soft 0.5-2 keV). 
We used them to the convert the hard X-ray luminosities (2-10 keV band) to observed fluxes in the soft and the hard bands.  

\subsection{Galactic extinction and number counts}
We include the effects of the Galactic foreground extinction following the map from \citet{HI4PI2016AA594A116H}. 
We use an average X-ray extinction law for AGN. 
It is computed by modelling analytically the spectra emerging through an absorber with an effective optical depth defined as the geometric mean between scattering and absorption \citep{Yaqoob97}; 
we use the absorption cross-sections of \citet{1983ApJ...270..119M}. 

\begin{figure*}
\centering
\includegraphics[width=0.68\columnwidth]{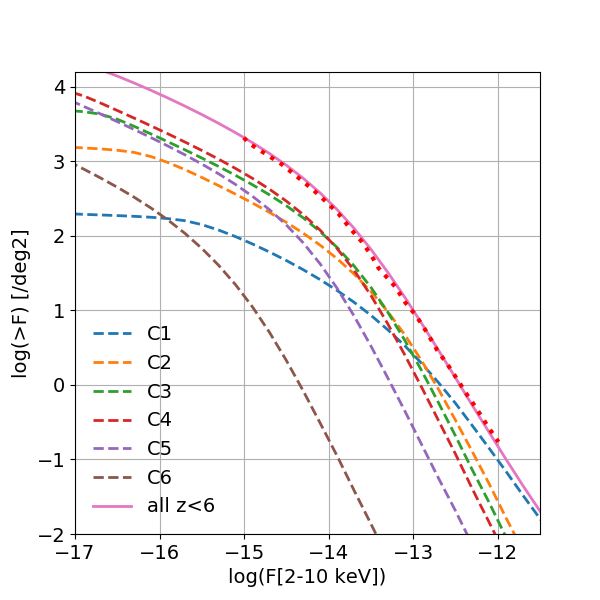}
\includegraphics[width=0.68\columnwidth]{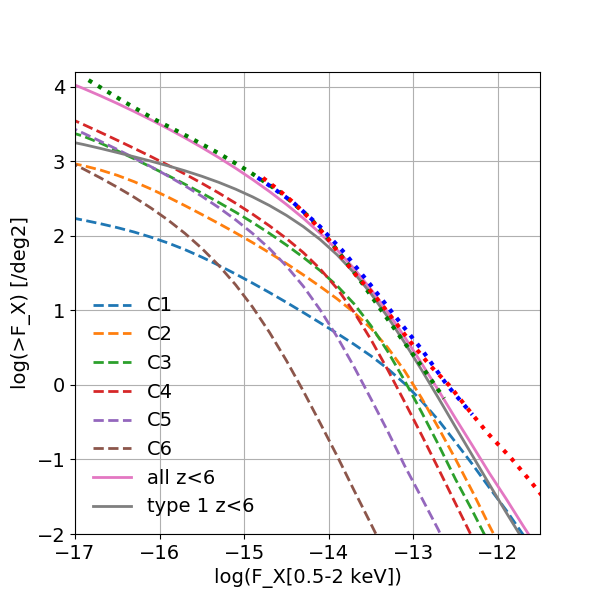}
\includegraphics[width=0.68\columnwidth]{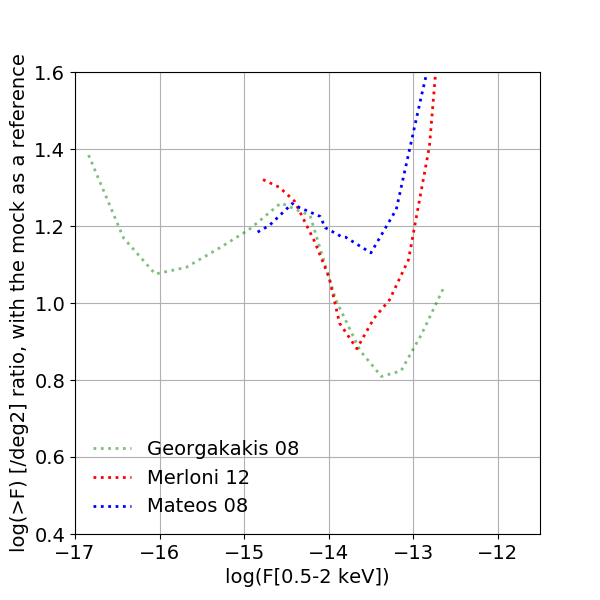}
\caption{\label{fig:lc:logN:logS:agn} logN-logS: AGN cumulative number density as a function of observed X-ray flux as in the mock catalogue between redshift 0 and 6 (pink solid line). 
The left (middle) panel shows the logN logS relation for the hard 2-10 keV band (soft, 0.5-2 keV band) X-ray flux, divided into 6 redshift bins corresponding to the light cone shells (dashed lines) with 
C1: $0<z<0.36$; 
C2: $0.36<z<0.77$; 
C3: $0.77<z<1.34$; 
C4: $1.34<z<2.19$; 
C5: $2.19<z<3.55$; 
C6: $3.55<z<6$.
This is compared to \citet{2008MNRAS.388.1205G,Mateos2008AA...492...51M,arxiv12093114_Merloni} (green, blue and red dotted lines).
In the right panel, is shown the ratio of these observed curves to the one from the mock for the soft X-ray flux. 
Around the flux limit of eROSITA, $\log_{10}(F^{0.5-2}_X) =-14$, the agreement is good, which is the key issue for the validity of the mock. 
At fainter fluxes, the mock is missing about 20\% of the sources. 
At the bright end because of the way the sampling is done and the finite volume of the simulation shells, we miss the brightest objects. 
The gray solid curve on the middle panel shows the logN-logS relation when restricted to type 1 AGNs.
}
\end{figure*}

To determine the number of AGN as a function of flux (logN-logS relation) \citet{2008MNRAS.388.1205G,Mateos2008AA...492...51M,arxiv12093114_Merloni}, 
we consider areas with Galactic latitude $|b_{gal}|>$20 degrees to avoid regions of high Galactic extinction. 
The logN-logS relation is shown for the hard (left panel) and soft X-ray flux (right panel) on Fig. \ref{fig:lc:logN:logS:agn}. 
The mock catalogue reproduces this to within $\pm20\%$, see Fig. \ref{fig:lc:logN:logS:agn} right panel. 
The shape of the relation varies in the literature at the 40\% level, but the total number of AGN with a flux greater than 
$F>10^{-14}\mathrm{erg cm}^{-2} \mathrm{s}^{-1}$ 
in the mock is within a few percent of the measurements from \citet{2008MNRAS.388.1205G,Mateos2008AA...492...51M,arxiv12093114_Merloni}, where the published values are in good agreement with each other. 
At the bright end, it is likely that the mock is missing objects due to the XLF sampling bias at high luminosities. 
At the faint end and at low redshift, we may be missing objects due to the resolution of the simulation. 

\subsection{eROSITA sensitivity and limiting fluxes}
eROSITA will scan the full sky every 6 months for a period of 4 years \citep{arxiv12093114_Merloni}, the data available after the first full sky scan is named `eRASS:1', after the second scan `eRASS:2' and so on until the final `eRASS:8', in order of increasing survey depth. 
We adopt here a preliminary eROSITA exposure map \citet{arxiv12093114_Merloni} which assumes a simple scanning strategy. 
The median and mean exposure times for eRASS:8 are 1721s and 2178s. The minimum is 804s and the maximum 57421s. 
We draw flux limits depending on the exposure time following \citet{2018AA...617A..92C}, Appendix A. 
For each object, we sample from the detection probability distribution function given at its sky location to determine if it is detected or not. This detection probability function is calibrated through detailed simulation of the measurement process  \citep[see][]{2018AA...617A..92C}.
We quote flux limits for 5$\sigma$ point source detection i.e. constrained by a spurious detection rate smaller than 1 in a million. 
For a 2ks exposure (typical of eRASS8 on the equator), the 50\% completeness flux is $1.1\times10^{-14} \mathrm{erg cm}^{-2} \mathrm{s}^{-1}$.
Decreasing the significance of the detection to $3\sigma$ increases the spurious detection rate to 1/370=0.2\% and decreases the flux limit by a factor of $\sim 2$. 
In this analysis, 
we consider three different catalogs, obtained with three different detection limits for three different phases of the survey: an "early" survey catalogue compiled from
5$\sigma$ detections after three full-sky scans (named eRASS3 hereafter), a "secure" 
5$\sigma$ detection catalogue after 8 full-sky scans (eRASS8), 
and 3$\sigma$ detection catalogue after 8 full-sky scans (SNR3). 
For the rest of the paper, unless mentioned otherwise, we adopt the  `eRASS8' as the baseline case (the other cases are presented in an Appendix). 

\subsection{Optical and X-ray AGN classification}
\label{subsec:classification}
In order to associate optical types to X-ray detected AGN in a meaningful and realistic way, we use the `obscuration-matrix' derived from \citet{2014MNRAS.437.3550M}. This describes, on the basis of the XMM-Newton X-ray selected AGN catalogue in the COSMOS field, the relative fraction of X-ray or optically obscured AGN by splitting them into four categories: unobscured in both optical and X-rays ('11'), X-ray obscured and optically unobscured ('12'), X-ray unobscured and optically obscured ('21'), obscured in both X-rays and optical ('22'). Here, optically "obscured" means that no broad line or blue accretion disc continuum (big blue bump) is detected in the optical spectrum.
We parameterize the boundaries as a function of X-ray luminosity between the 21 and the 11 as well as between the 22 and 12 classes (visible on Fig. 12 of \citealt{2014MNRAS.437.3550M}) with a Gauss error function ($\epsilon$ defined in Eq. \ref{eqn:gauss:err:fct}):
\begin{equation}
fr_{\rm split}(l_x) = \left[\frac{1}{2} + \frac{1}{2} \epsilon\left(\frac{44-l_x}{0.9}\right)\right]0.69+0.26,
\end{equation}
where $l_x$ and $\epsilon$ are defined in Eq. (\ref{eq:def:lx:logLX}) and (\ref{eqn:gauss:err:fct}).
This boundary further splits the X-ray obscured-unobscured classification. 

\begin{figure}
\includegraphics[width=.85\columnwidth,type=png,ext=.png,read=.png]{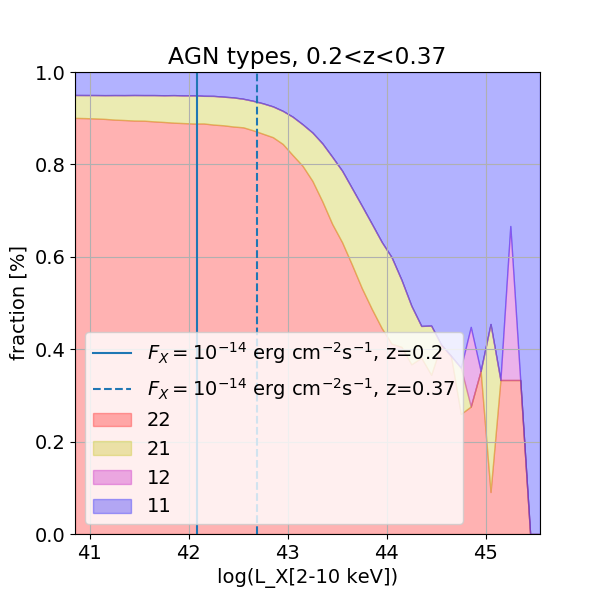}
\includegraphics[width=.85\columnwidth,type=png,ext=.png,read=.png]{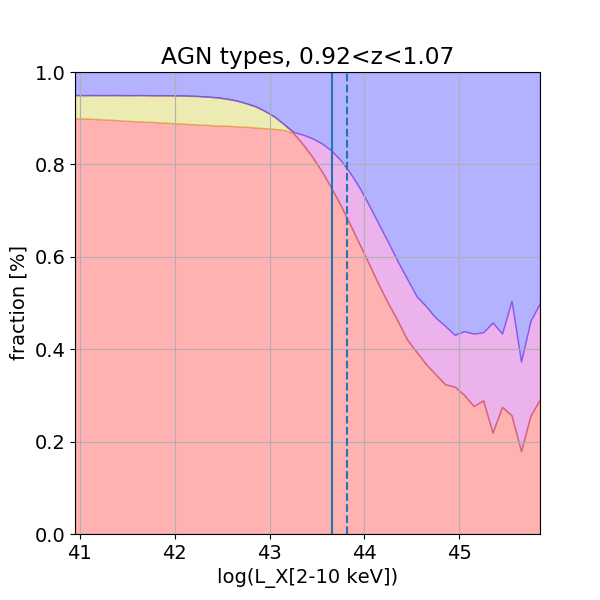}
\includegraphics[width=.85\columnwidth,type=png,ext=.png,read=.png]{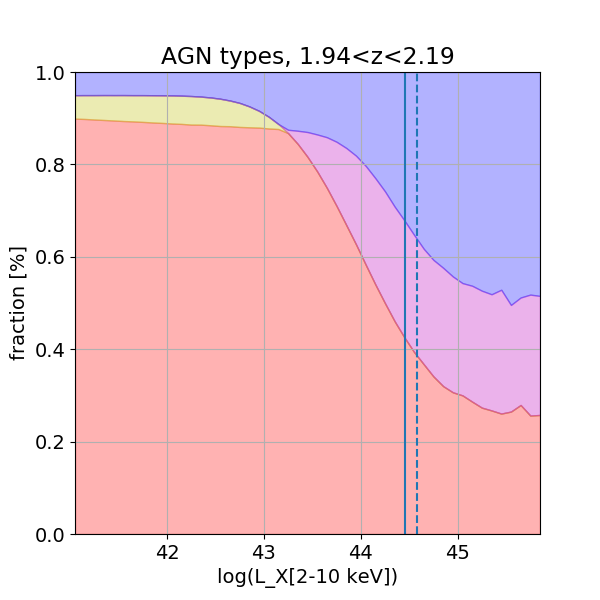}
\caption{Fractions of AGNs as a function of hard X-ray luminosity (2-10 keV band) in each category: unobscured in both optical and X-rays ('11'), X-ray obscured and optically unobscured ('12'), X-ray unobscured and optically obscured ('21'), obscured in both X-rays and optical ('22'). The vertical lines show the luminosity corresponding to the flux of $10^{-14}\mathrm{erg cm}^{-2} \mathrm{s}^{-1}$ (close to the eROSITA detection limit) at the boundaries of the redshift interval considered.
}
\label{fig:obscured:fractions}
\end{figure}

We compute the obscured fraction of AGN ($fr_o$) as a function of X-ray luminosity using the boundary $\log_{10}(n_H)>22$. 
$fr_o-fr_{\rm split}$ gives the fraction of optically unobscured AGN among the X-ray obscured population: type 12. 
We select them randomly among the X-ray obscured ($\log_{10}(n_H)>22$) sources to match the fraction needed. 
$fr_{\rm split}-fr_o$ gives then the fraction of optically obscured AGN among the X-ray unobscured: type 21. 
We select these randomly among the X-ray unobscured ($\log_{10}(n_H)<22$) to match the fraction needed. 
The X-ray obscured (unobscured) AGN that were not assigned to type 12 (21) are assigned type 22 (11). 
Among the X-ray obscured AGNs ($n_H>22$), most become type 22 and some type 12 (at high luminosity).
Figure \ref{fig:obscured:fractions} shows the classification obtained as a function of $L_X$ at three redshifts: 0.3, 1, 2. 

\subsection{Spectral energy distribution in the Ultra-Violet, Optical and Infrared}
Determining redshifts for the eROSITA AGN is key for most scientific applications. 
A redshift measurement with an accuracy of $\sim1\%$ and outlier fraction of 5\% is possible using a set of broad band photometric magnitudes \citep[i.e. `photo-z'][]{2011ApJ...742...61S,2018NatAs.tmp...68S,2018MNRAS.473.4937S}. 
Photo-z of this accuracy are sufficient for luminosity function studies, for example, and the SEDs more generally can provide information on classification. 
A redshift measurement with accuracy 0.05\% and an outlier fraction smaller than 1\% is feasible using spectroscopy \citep[i.e. `spec-z'][]{2015ApJS..221...27M,2016AJ....151...44D,2017MNRAS.469.1065D}. 
Spec-z allow studies of the clustering and of spectral line properties, including more accurate and details (sub)classifications. 
To emulate photo-z and spec-z, we create an empirical description of AGNs in the UV, optical and IR.

In Sec. \ref{subsec:sdss:r:band}, we compute the SDSS $r$-band magnitude for each AGN. 
Then in Sec. \ref{subsec:SED:broad}, we link each AGN to a spectral energy distribution extending from the UV to the IR to compute a set of broad band magnitudes. 
Finally in Sec. \ref{subsec:spectra}, we link each AGN to a medium resolution optical spectrum to emulate spectroscopic observations. 

\subsubsection{Optical SDSS $r$-band magnitude}
\label{subsec:sdss:r:band}
In this section, we describe the method we use to compute the SDSS $r$-band magnitudes \citep{1996AJ....111.1748F}.
To do this, we used the observed $F_{X}/F{opt}$ ratio or more precisely the relation between soft X-ray flux and $r$-band magnitude from existing surveys to calibrate the relationship and its scatter. The $r$ band magnitude can then also be used to normalise the overall SED. The $F_{X}/F{opt}$ relationship could in principle depend on the optical type, because of the differential effects of obscuration on the X-ray and optical fluxes, and the differing contributions of the host galaxy. 

We therefore examine this relation for three optical classes of AGNs: 
\begin{itemize}
\item Optical spectroscopic type 1 corresponding to type 11 or type 12, with quasar-like spectra. 
\item Optical spectroscopic type 2 corresponding to type 22 or type 21, with type2-like spectra.
\item The `elusive' AGN, or XBONGS (X-ray bright, optical normal galaxies) \citep{Comastri2002,Menzel2016}; we call these type 3 AGN. We take a random 20\% of the optical spectroscopic type 2 AGNs to be represented by this class. Their spectrum is galaxy-like.
\end{itemize}

\begin{figure*}
\includegraphics[width=1.3\columnwidth,type=png,ext=.png,read=.png]{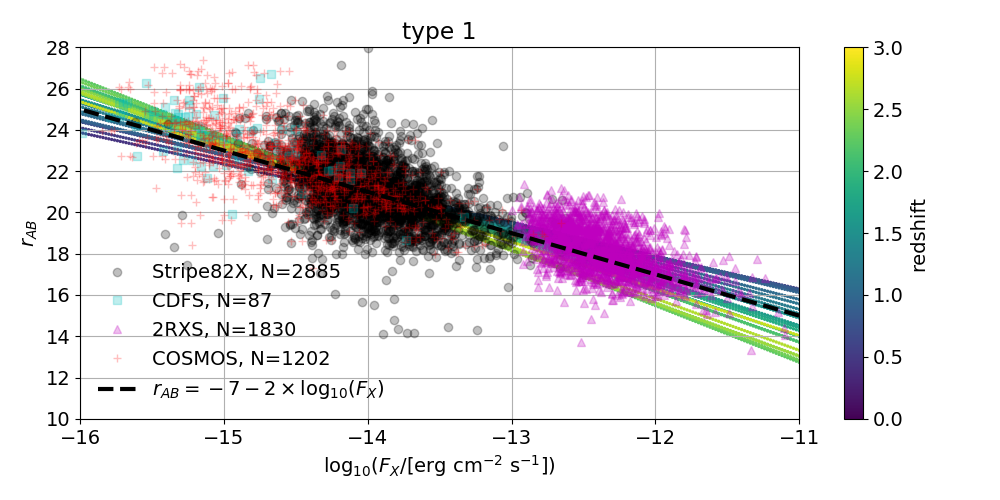}
\includegraphics[width=.68\columnwidth,type=png,ext=.png,read=.png]{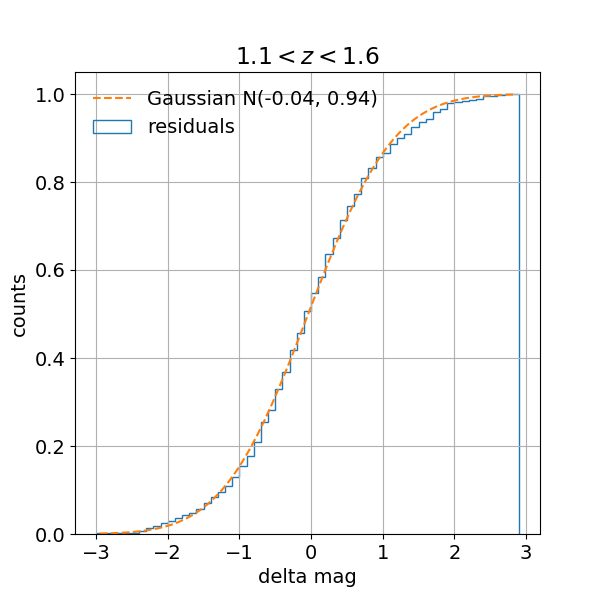}
\includegraphics[width=1.3\columnwidth,type=png,ext=.png,read=.png]{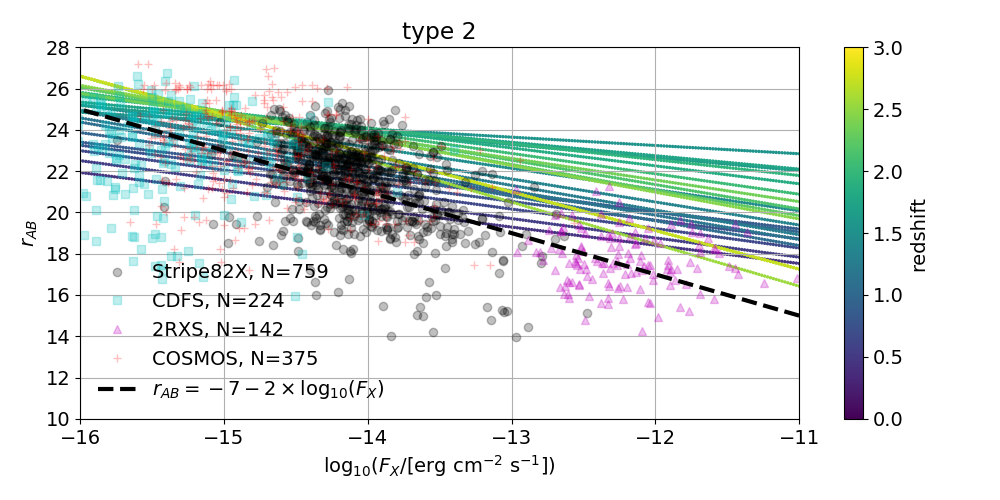}
\includegraphics[width=.68\columnwidth,type=png,ext=.png,read=.png]{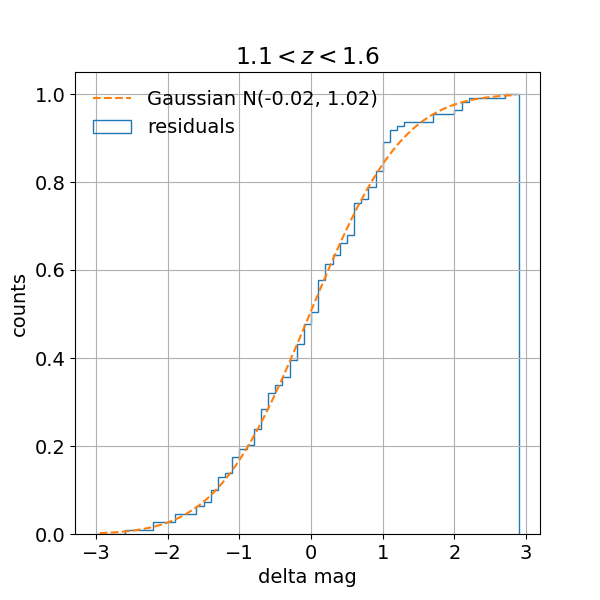}
\includegraphics[width=1.3\columnwidth,type=png,ext=.png,read=.png]{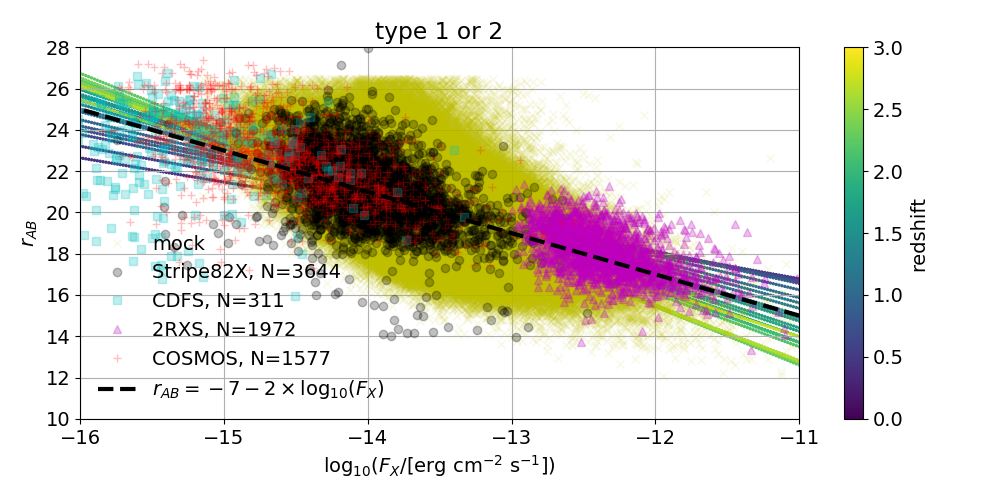}
\includegraphics[width=.68\columnwidth,type=png,ext=.png,read=.png]{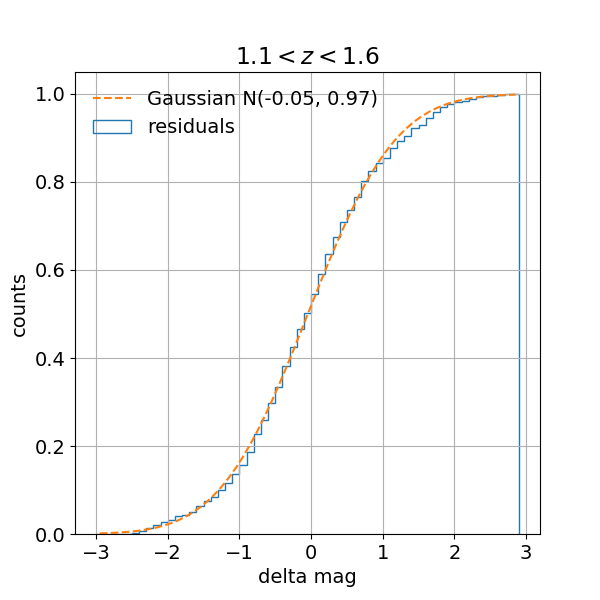}
\caption{$r$ magnitude vs. soft X-ray flux (left) and cumulative distribution of residuals around the model (right) for type 1 AGN (top), type 2 AGN (middle), and both (bottom) in the redshift bin $1.1<z<1.6$.
The CDFS, Stripe82X, COSMOS and 2RXs catalogues are shown in cyan circles, black circles, red crosses, magenta triangles \citep{2014ApJ...796...60H,2016MNRAS.456.1359F,2017ApJ...850...66A,2016ApJ...817...34M,2018MNRAS.473.4937S}. 
The yellow crosses in the bottom left panel show the eRASS8 mock.
The colored lines show the best fits as a function of redshift. 
The black dashed line shows the relation adopted. 
}
\label{fig:rmag:FX}
\end{figure*}

We performed a fit to the ratio of the 0.5-2 keV flux to the optical r band magnitude, using the CDFS, Stripe82X, COSMOS and 2RXs catalogues \citep{2014ApJ...796...60H,2016MNRAS.456.1359F,2017ApJ...850...66A,2016ApJ...817...34M,2018MNRAS.473.4937S}. 
In these catalogs, the AGN are sorted according to optical type 1 (unobscured), 2 (obscured) or 3 (galaxy-like), as defined by the best-fitting SED template or spectroscopic information. 
Note that the classification is not perfect and the three classes are to some extent contaminating each other. 
Also the total number of type 2 and 3 is small with respect to the total number of type 1. 
So we fit one relation for type 1 and one for type 2 and 3 together. 
We find that a mean relation works for all populations: 
\begin{equation}
\bar{r}(\log_{10}F_X) = -2\log_{10}(F_X) - 7
\end{equation}
The scatter around the relation is well approximated by a Gaussian with unity scatter for all populations:  
\begin{align}
\sigma = \mathcal{N}(0.0, 1.0) 
\end{align}
Fig. \ref{fig:rmag:FX} shows the data considered in the $r$ magnitude vs. soft X-ray flux plane. 
It shows that a single relation accounts well for all types. 
It shows a hint of trend with redshift, more data is required to confirm this trend. 

Current surveys (\textit{e.g.} COSMOS, Stripe82, CDFS) lack sufficient number of AGN to constrain an eventual joint fit of the X-ray and optical luminosity functions.

Note that current data does not contain sufficient number of AGN to constrain an eventual joint fit of the X-ray and optical luminosity functions. 
A complete discussion on the possible universality of this relation is left for future studies. 
Indeed, future surveys such as eROSITA, 4MOST and the Maunakea Spectroscopic Explorer \citep{2019Msngr.175...42M,2019Msngr.175...50R,2019arXiv190404907T} will hopefully settle this. 	

\subsubsection{Spectral energy distribution and broad band photometry}
\label{subsec:SED:broad}
\begin{figure}
\centering
\includegraphics[width=8cm]{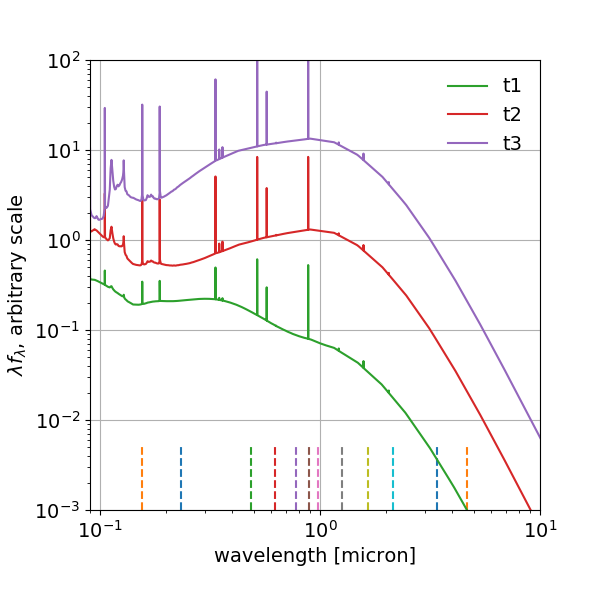}
\caption{Flux v.s. wavelength (microns) for the three spectral energy distributions used to compute AGN broad band magnitudes. 
The curves t1, t2, t3 represent the optical classes described in the text: type 1, 2, 3.
The vertical dashed lines show the mean wavelengths of the broad-band filters used to compute magnitudes, from left to right:
GALEX-NUV,
GALEX-FUV,
HSC-g,
HSC-r,
HSC-i,
HSC-z,
HSC-y,
UKIDSS-J,
UKIDSS-H,
UKIDSS-K,
WISE-W1,
WISE-W2.
}
\label{fig:SED}
\end{figure}

To test the algorithms that identify the AGN counterparts in the UV, optical and IR and measure the photo-z, we need to generate mock magnitudes corresponding to those that will be available for the datasets which will be used in reality. 
We use spectral energy distributions (SED) models from \citet{ciesla2015AA576A10C} to deduce UV, optical and near IR magnitudes for all AGN simulated in the mock. 
The SEDs are shown in Fig. \ref{fig:SED}. 
We use a model SED for each spectroscopic optical type: 1,2,3.
The following band passes are simulated: GALEX-NUV, GALEX-FUV, HSC-g, HSC-r, HSC-i, HSC-z, HSC-y, UKIDSS-J, UKIDSS-H, UKIDSS-K, WISE-W1, WISE-W2 \citep{GALEX2005ApJ619L1M,HSC_filters_2018PASJ7066K,Hewett2006MNRAS367454H,wright_2010}.
These surveys are of particular interest because they are located in the performance verification equatorial commissioning field of eROSITA. 
We use the model SED, rescaled to the SDSS $r$-band magnitude, and create a magnitude array and its uncertainty to mimic the different depths of the GALEX, HSC, VIKING and WISE surveys \citep{GALEX2007ApJS173682M,HSC_DR1_2018PASJ70S8A,viking2016yCat23430E,DECALS_2018arXiv180408657D}. 
To mimic the depth, we fit with a polynomial of second or third degree the magnitude -- error relation in each of these surveys in a representative region. Table \ref{tab:mag:err} gives the coefficients of the polynomial 
\begin{equation}
\label{eqn:mag:err}
 \log_{10}(\sigma_m(m))=\Sigma_i c_i m^i
\end{equation}
that links an AB magnitude, $m$, to its uncertainty $\sigma_m$. 
For individual AGNs, we draw uncertainties around this relation with a Gaussian scatter of 0.158 dex. 

\begin{table}
	\centering
	\caption{\label{tab:mag:err}Polynomial coefficients of the magnitude error relation given in Eq. \ref{eqn:mag:err}. 
	 $\bar{\lambda}$ is the mean wavelength of the broad band filter in microns. 
	}	
	\begin{tabular}{cc rr rrr rrr rrrrrrrrrrr} 
	\hline   
\hline
band & $\bar{\lambda}$ & \multicolumn{3}{c}{polynomial coefficients}  \\ 
     & $c_2$ & $c_1$ & $c_0$ \\ \hline
GALEX-NUV &0.234& 7.24e-03 & -4.28e-04 & -5.13 \\ 
GALEX-FUV &0.155& 6.39e-03 & -2.43e-04 & -4.67 \\  
HSC-g     &0.485&& 0.337  & -9.727 &                 \\
HSC-r     &0.627&& 0.346  & -9.788 &                 \\
HSC-i     &0.777&& 0.353  & -9.905 &                 \\
HSC-z     &0.892&& 0.353  & -9.646 &                 \\
HSC-y     &0.979&& 0.341  & -9.075 &                 \\
UKIDSS-J  &1.256& 0.009 & -0.010  & -4.611    \\
UKIDSS-H  &1.651& 0.012 & -0.091  & -3.910    \\
UKIDSS-K  &2.153& 0.014  & -0.124 & -3.521    \\
WISE-W1   &3.4&& 0.394 & -9.129                   \\
WISE-W2   &4.652&& 0.396 & -8.797                   \\
\hline 
\end{tabular}
\end{table}

\subsubsection{Optical medium resolution spectra}
\label{subsec:spectra}
We link each AGN in the mock to a representative medium resolution (R$\sim$3000) optical spectrum by stacking SDSS spectra \citep{Paris2018AA613A51P,SDSS_DR14_2018ApJ23542A} with the archetype method \citet{Zhu_2016arXiv160607156Z} to represent each of the three types as a function of redshift. 

We create a set of medium resolution optical spectra for the type 1, 2, 3 AGNs as well as for red, elliptical galaxies and for blue, star-forming galaxies. 

For the type 1 AGNs, there exist a large number of optical spectra, and the stacking method permits the construction of high-quality stacks that are almost noise free, see Fig \ref{fig:spectra:qso}. 

For the type 2 AGNs, the number of available spectra is small, so that the stacks have noise, in particular at redshift higher than 1, see Fig \ref{fig:spectra:agn}. 

For the type 3 AGNs, we use the elliptical, red galaxy spectrum model, see Fig \ref{fig:spectra:lrg}. 

To construct galaxy archetypes we use SDSS spectra as selected in \citep{2009ApJS..182..543A,2013AJ....145...10D,2016ApJS..224...34P} for the elliptical galaxies (Fig \ref{fig:spectra:lrg}) and as in \citep{2013MNRAS.428.1498C,2016A&A...592A.121C,2017MNRAS.471.3955R} for the star-forming galaxies (Fig \ref{fig:spectra:elg}). 

We use the Planck dust maps with the \textsc{dustmap} package\footnote{\url{http://dustmaps.readthedocs.io/en/latest/index.html}} to redden the optical SDSS $r$-band magnitude as well as the spectra. 
We use the \textsc{extinction}\footnote{\url{https://extinction.readthedocs.io}} package to attenuate magnitudes and spectra with the \citet{1999PASP..111...63F} extinction law (The $r$ LF comparison discussed above is done on de-reddened magnitudes). 
This is used below for spectroscopic exposure time calculation. 

%%%%%%%%%%%%%%%%%%%%%%%%%%%%%%%%%%%%%%%%%%%%%%%%%%%%%%%%%%%%%%%%%%%%%%%%%%%%%%%%%%%%%%
%%%%%%%%%%%%%%%%%%%%%%%%%%%%%%%%%%%%%%%%%%%%%%%%%%%%%%%%%%%%%%%%%%%%%%%%%%%%%%%%%%%%%%
%%%%%%%%%%%%%%%%%%%%%%%%%%%%%%%%%%%%%%%%%%%%%%%%%%%%%%%%%%%%%%%%%%%%%%%%%%%%%%%%%%%%%%
% RESULTS
%%%%%%%%%%%%%%%%%%%%%%%%%%%%%%%%%%%%%%%%%%%%%%%%%%%%%%%%%%%%%%%%%%%%%%%%%%%%%%%%%%%%%%
%%%%%%%%%%%%%%%%%%%%%%%%%%%%%%%%%%%%%%%%%%%%%%%%%%%%%%%%%%%%%%%%%%%%%%%%%%%%%%%%%%%%%%
%%%%%%%%%%%%%%%%%%%%%%%%%%%%%%%%%%%%%%%%%%%%%%%%%%%%%%%%%%%%%%%%%%%%%%%%%%%%%%%%%%%%%%

\section{Results: the predicted eROSITA AGN population}
\label{sec:results}

\begin{figure*}
\centering
\includegraphics[width=17cm]{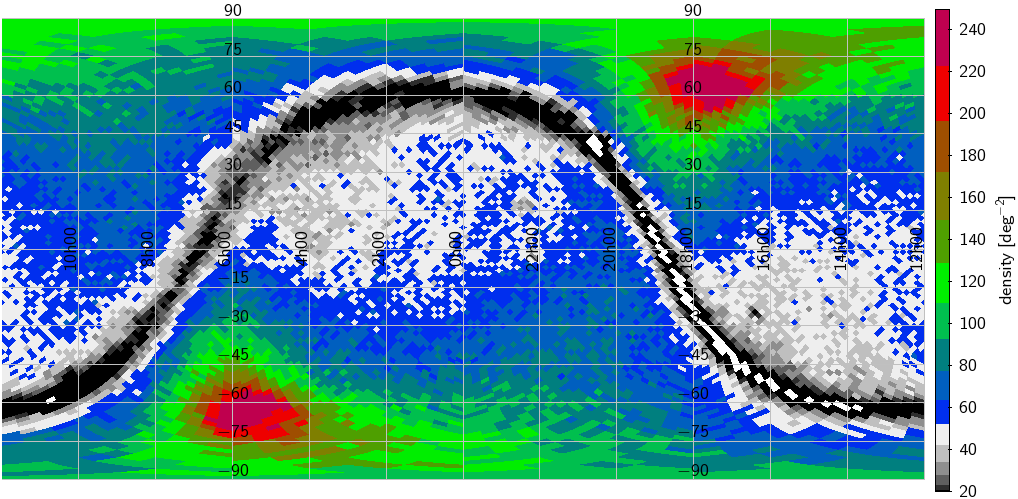}
\caption{Distribution on the sky of the expected number density of X-ray AGNs after the fiducial eRASS8 observation plan i.e. after exposure map and Galactic extinction is taken into account (number counts averaged on 3.35 deg$^2$ pixels). %\textcolor{blue}{Add ecliptic equator + poles?}
}
\label{fig:sky:distributions}
\end{figure*}

The main application of the model is the prediction of the distribution of eROSITA AGN on the sky, in redshift and magnitude. 
These distributions show how the eROSITA all-sky survey will sample our past light cone in an unprecedented fashion. 

The total expected number of AGNs is 2.3 (0.9, 3.5) million for the eRASS8 scenario (eRASS3, SNR3); previous estimates were around 3 for eRASS8. 
The discrepancy is mainly due to previous sensitivity maps \citep{arxiv12093114_Merloni,2013AA558A89K} that were more optimistic than the more recent one from \citet{2018A&A...617A..92C}. 
As the instrumental background for eROSITA at the L2 point is highly uncertain, the exact sensitivity as a function of exposure time is not easy to predict. 
Having the three scenarios is thus handy to see how total number evolve with total exposure time assumed (3/8 exposure time ratio between eRASS3 and eRASS8 scenarios) and detection significance (3$\sigma$ and 5$\sigma$ detection in the SNR3 and eRASS8 scenarios).

The expected variation in AGN number density as a function of right ascension and declination in a `car' projection is shown on Fig. \ref{fig:sky:distributions}  (Fig. \ref{fig:sky:distributions:2} shows the same projection for the eRASS3 and SNR3 scenarios). 
Away from the poles and the milky way, eROSITA will typically detect between 60 and 100 AGN per square degree.  
The ecliptic poles, where the eROSITA exposure is the longest, are the most densely populated regions with densities around 250 deg$^{-2}$ in the eRASS8 scenario. 
For the eROSITA telescopes, the angular half energy width at 1.5 keV is $\approx$16 arc seconds on-axis, and degrades to an arc minute off-axis. 
Source confusion thus influences at source densities above 1 arc minute$^{-2}\sim1100$ deg$^{-2}$; thus,
source confusion can safely be ignored for most areas to be surveyed by eROSITA. 
Note that the maximum density (in the extra-Galactic sky) is reached in the SNR3 scenario with a maximum of 400 deg$^{-2}$ in the poles, so that in this specific case and region source confusion may have an impact.

\begin{figure}
\centering
\includegraphics[width=\columnwidth,type=png,ext=.png,read=.png]{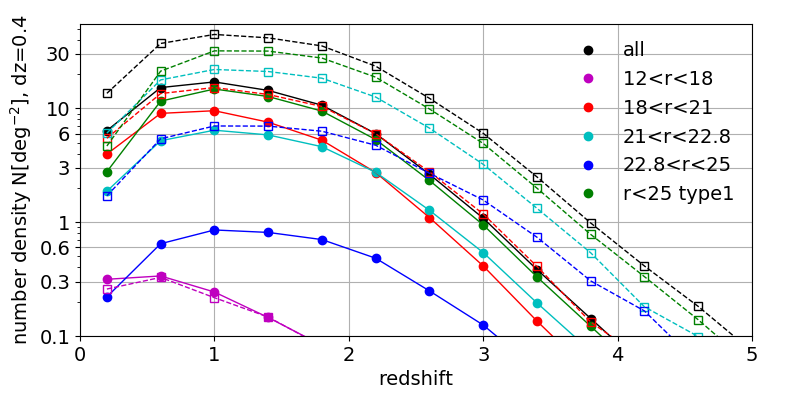}
\caption{\label{fig:redshift:distributions:eRASS8}Predicted redshift distributions per deg$^2$ (intervals of $dz=0.4$) for the fiducial eRASS8 scenario and for different magnitude cuts in the SDSS $r$-band: $12<r<18$, $18<r<21$, $21<r<22.8$, $22.8<r<25$, all, corresponding to Table \ref{tab:AGN:NZ:ALL:eRASS8}. 
The extra-Galactic distribution is represented with filled circles and the poles with empty squares. 
This was computed over an area of 27,143 (626.7) deg$^{2}$ for the extra-Galactic sky (poles), namely $|b_{gal}|>20^\circ$ ($|b_{ecl}|>80^\circ$). 
}
\end{figure}

We give the average redshift distribution of the AGNs for scenarios eRASS8, eRASS3, SNR3 in the Tables \ref{tab:AGN:NZ:ALL:eRASS8}, \ref{tab:AGN:NZ:ALL:eRASS3},  \ref{tab:AGN:NZ:ALL:SNR3}. 
They give the average redshift distributions over the extra-Galactic sky (27k deg$^2$) and at the poles. 
We predict an average detection rate of $\sim74$ AGN deg$^{-2}$ (218) on the extra-Galactic sky (pole). 
We find good agreement with previous predictions concerning the shape of the redshift distribution \citep{arxiv12093114_Merloni,2013AA558A89K}. 
The redshift distribution is illustrated in Figs. \ref{fig:redshift:distributions:eRASS8}, \ref{fig:redshift:distributions:eRASS3}, \ref{fig:redshift:distributions:SNR3} for the scenarios eRASS8, eRASS3, SNR3.
These show how the poles compare to the general extra-Galactic sky area.
The redshift distributions are shown in four magnitude ranges that are appropriate for the facilities that will be used for optical follow-up, see Sec. \ref{subsection:specz:follow:up}.

\begin{table*}
	\centering
	\caption{\label{tab:AGN:NZ:ALL:eRASS8}Number density of AGN as a function of redshift for the eRASS8 scenario for different optical $r$-band magnitude cuts ($12<r<18$, $18<r<21$, $21<r<22.8$, $22.8<r<25$, type 1 $r<25$,  all $r<25$). 
	The first two columns give the boundaries of each redshift bins. 
	All other columns give $\frac{dN_{AGN}}{deg^2}$ the average number of AGN per square degree in that redshift bin for the extra-Galactic sky (x) and for the poles (p). 
	This was computed on an area of 27,143 (626.7) deg$^{2}$ for the extra-Galactic sky (poles), namely $|b_{gal}|>20^\circ$ ($|b_{ecl}|>80^\circ$). 
	}	
	\begin{tabular}{cc r r rrr rrr rrrrrrrrrrr} 
	\hline   
\hline
\multicolumn{2}{c}{$z$ bin}  &
\multicolumn{2}{c}{$12<r<18$} & 
\multicolumn{2}{c}{$18<r<21$} & 
\multicolumn{2}{c}{$21<r<22.8$} & 
\multicolumn{2}{c}{$22.8<r<25$} &
\multicolumn{2}{c}{type 1} &
\multicolumn{2}{c}{all} \\ 

min & max &
 x & p &
 x & p &
 x & p &
 x & p & 
 x & p &
 x & p \\

\hline
    0.0 &     0.4 &     0.3148 &     0.2585 &     3.9610 &     5.4617 &     1.8695 &     6.0537 &     0.2206 &     1.7041 &     2.7738 &     4.6623 &     6.3681 &    13.5338 \\
    0.4 &     0.8 &     0.3368 &     0.3271 &     9.0241 &    13.4732 &     5.1919 &    17.7606 &     0.6465 &     5.3947 &    11.5537 &    21.2087 &    15.2055 &    37.1024 \\
    0.8 &     1.2 &     0.2434 &     0.2170 &     9.5206 &    15.2156 &     6.4195 &    21.9283 &     0.8538 &     6.9935 &    14.7708 &    31.9502 &    17.0446 &    44.5315 \\
    1.2 &     1.6 &     0.1455 &     0.1468 &     7.5546 &    13.2227 &     5.8378 &    21.0427 &     0.8124 &     6.9743 &    12.6885 &    31.7970 &    14.3571 &    41.5717 \\
    1.6 &     2.0 &     0.0828 &     0.0814 &     5.2457 &    10.3730 &     4.5988 &    18.3749 &     0.7029 &     6.2898 &     9.4502 &    27.6230 &    10.6380 &    35.2978 \\
    2.0 &     2.4 &     0.0364 &     0.0399 &     2.7281 &     5.9739 &     2.7400 &    12.5573 &     0.4821 &     4.7373 &     5.3163 &    18.7626 &     5.9915 &    23.4393 \\
    2.4 &     2.8 &     0.0120 &     0.0112 &     1.0968 &     2.7668 &     1.2710 &     6.6983 &     0.2490 &     2.6886 &     2.3244 &     9.8688 &     2.6315 &    12.2398 \\
    2.8 &     3.2 &     0.0034 &     0.0000 &     0.4124 &     1.1791 &     0.5407 &     3.2199 &     0.1248 &     1.5653 &     0.9459 &     4.9112 &     1.0836 &     6.0282 \\
    3.2 &     3.6 &     0.0014 &     0.0016 &     0.1353 &     0.4085 &     0.1953 &     1.3227 &     0.0505 &     0.7372 &     0.3305 &     1.9881 &     0.3836 &     2.4971 \\
    3.6 &     4.0 &     0.0003 &     0.0000 &     0.0472 &     0.1324 &     0.0736 &     0.5377 &     0.0210 &     0.3048 &     0.1217 &     0.7802 &     0.1424 &     0.9829 \\
    4.0 &     4.4 &     0.0002 &     0.0016 &     0.0174 &     0.0622 &     0.0256 &     0.1803 &     0.0097 &     0.1659 &     0.0445 &     0.3319 &     0.0530 &     0.4117 \\
    4.4 &     4.8 &     0.0001 &     0.0000 &     0.0067 &     0.0239 &     0.0100 &     0.0989 &     0.0035 &     0.0574 &     0.0169 &     0.1388 &     0.0203 &     0.1819 \\
    4.8 &     5.2 &     0.0000 &     0.0000 &     0.0028 &     0.0207 &     0.0046 &     0.0335 &     0.0010 &     0.0191 &     0.0066 &     0.0543 &     0.0084 &     0.0734 \\
    5.2 &     5.6 &     0.0000 &     0.0000 &     0.0008 &     0.0032 &     0.0018 &     0.0191 &     0.0006 &     0.0096 &     0.0026 &     0.0271 &     0.0033 &     0.0319 \\
    5.6 &     6.0 &     0.0000 &     0.0000 &     0.0004 &     0.0000 &     0.0009 &     0.0064 &     0.0003 &     0.0080 &     0.0014 &     0.0128 &     0.0016 &     0.0144 \\ \hline
    0.0 &     6.0 &     1.1771 &     1.0850 &    39.7540 &    68.3170 &    28.7810 &   109.8344 &     4.1788 &    37.6497 &    60.3478 &   154.1170 &    73.9326 &   217.9376 \\
\hline 
\end{tabular}
\end{table*}

\subsection{Optical luminosity function}
\begin{figure*}
\centering
\includegraphics[width=.68\columnwidth,type=png,ext=.png,read=.png]{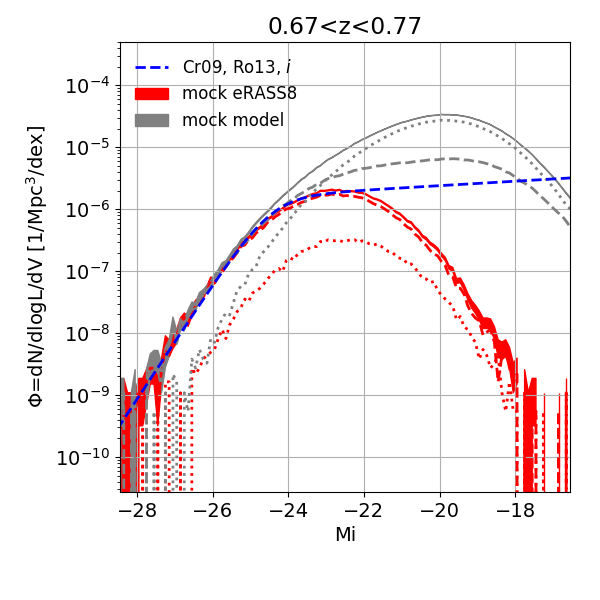}
\includegraphics[width=.68\columnwidth,type=png,ext=.png,read=.png]{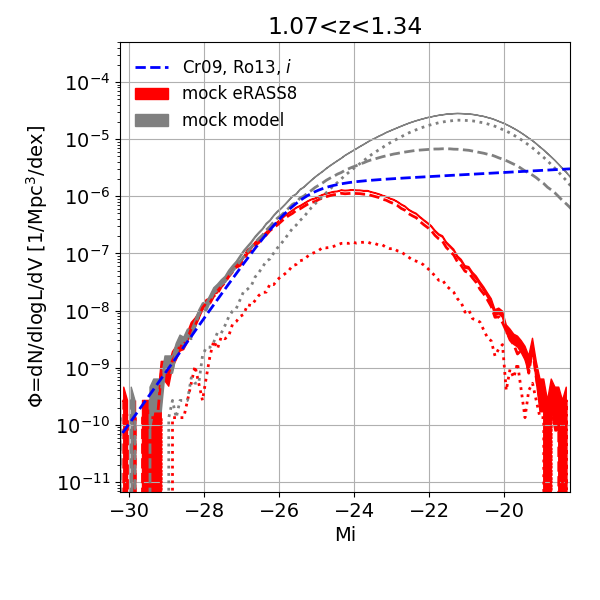}
\includegraphics[width=.68\columnwidth,type=png,ext=.png,read=.png]{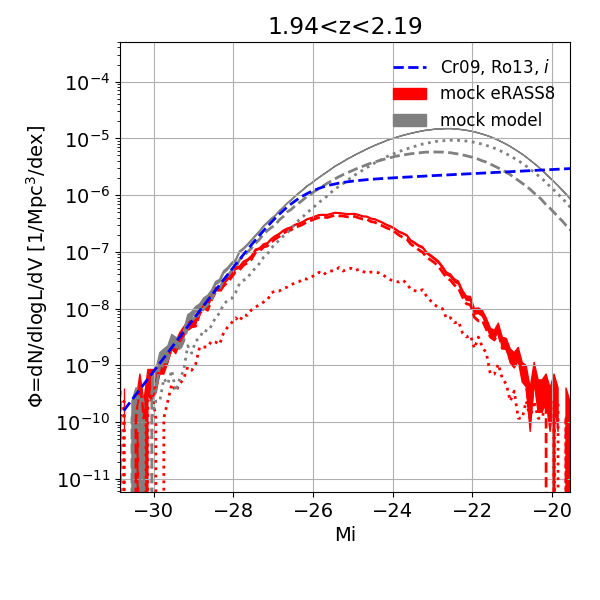}
\includegraphics[width=.68\columnwidth,type=png,ext=.png,read=.png]{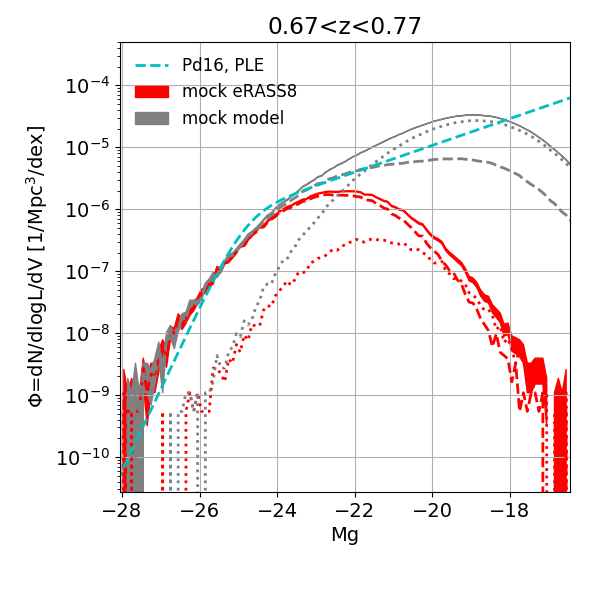}
\includegraphics[width=.68\columnwidth,type=png,ext=.png,read=.png]{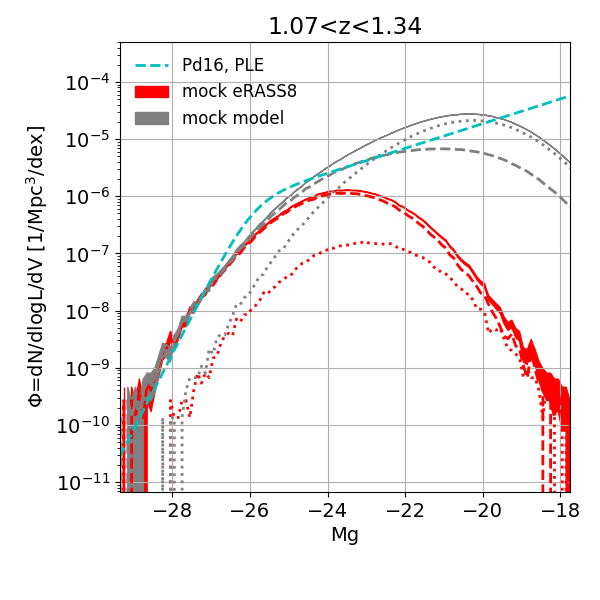}
\includegraphics[width=.68\columnwidth,type=png,ext=.png,read=.png]{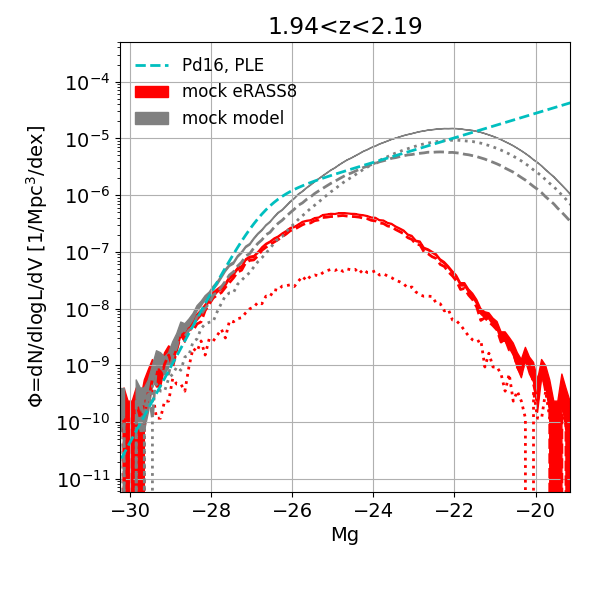}
\caption{Observed-frame optical luminosity functions of AGNs (color-shaded area) subdivided into optical type 1 AGNs (dashed) and type 2 and 3 AGNs (dotted). 
The full model, down to the resolution of the simulation is shown in gray and the eRASS8 mock in red. 
It is compared to the models based on observations from \citet{Croton2009MNRAS.394.1109C,Ross_N_2013ApJ77314R} (Cr09, Ro13) for the $i$-band (top row) and to the \citet{Caditz2017AA608A64C,Palanque2016AA587A41P} model for the $g$-band (bottom row). 
The agreement at the bright end is excellent. 
}
\label{fig:RLF}
\end{figure*}

\begin{figure}
\centering
\includegraphics[width=.85\columnwidth]{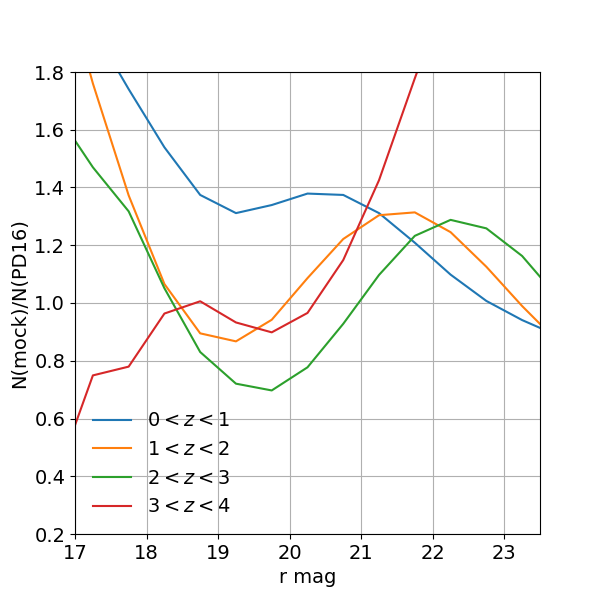}
\caption{Ratio of the number of optical type 1 AGNs predicted as a function of $r$-band magnitude in wide redshift bins by the mock and by the \citet{Caditz2017AA608A64C,Palanque2016AA587A41P} models (Table 7). 
In the range $0<z<3$, $19<r<23$, the mock is within $\pm40\%$ of the models. Outside of this range, the discrepancy can reach a factor of 2.
}
\label{fig:RLF:ratio}
\end{figure}

\begin{figure}
\centering
\includegraphics[width=8cm]{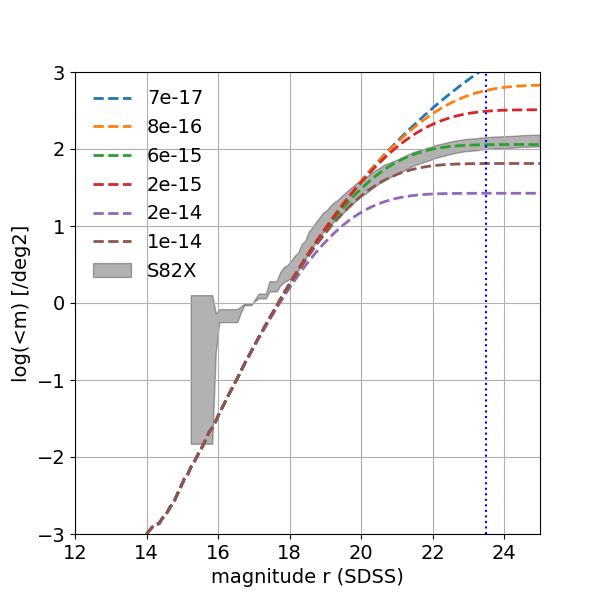}
\caption{logN-logR relation obtained from the mock catalogue compared to the relation obtained with the Stripe82X catalogue from \citet{2017ApJ...850...66A} (gray shaded area). For the mock, we show many lines corresponding to different X-ray flux limits $F^{0.5-2keV}_X > 10^{-14}, 2\times10^{-14}, 2\times10^{-15}, 6\times10^{-15}, 8\times10^{-16}, 7\times10^{-17}\mathrm{erg cm}^{-2} \mathrm{s}^{-1}$. }
\label{fig:logNlogR}
\end{figure}

With the SEDs described above, we obtain a set of optical AGN LFs. 
They have a double power-law shape like the XLF. 
As shown in Fig. \ref{fig:RLF}, our predicted optical LF agrees within a factor of 2 with the $i$-band LF of quasars \citep{Croton2009MNRAS.394.1109C,Ross_N_2013ApJ77314R} and with the $g$-band LF of a complete census of optically-selected AGN \citep{Palanque2016AA587A41P,Caditz2017AA608A64C} (PLE model). 
The observer-frame $g$ and $i$-band luminosity functions are shown in Fig. \ref{fig:RLF}. 
We show the complete model, limited by the resolution of the simulation (gray), and the eRASS8 mock (red), limited by the eROSITA selection function. 
We compare the predicted optical type 1 AGN number counts as a function of the observed $r$-band magnitude in four broad redshifts bins ($0<z<1$, $1<z<2$, $2<z<3$, $3<z<4$) with the observations from \citet{Palanque2016AA587A41P,Caditz2017AA608A64C} (Table 7). 
The number of AGN predicted by the mock is within $\pm40\%$ of the model prediction in the range $0<z<3$, $19<r<23$, see Fig. \ref{fig:RLF:ratio}. Outside of this range, the discrepancy can reach a factor of 2.
Finally, Fig. \ref{fig:logNlogR} shows the cumulative number density projected on the sky (logN-logR) of a set of X-ray selected AGN samples ($F^{0.5-2}_X>10^{-14}$, $2\times 10^{-14}$, $2\times 10^{-15}$, $6\times 10^{-15}$, $8\times 10^{-16}$, $7\times 10^{-17}\mathrm{erg cm}^{-2} \mathrm{s}^{-1}$) as obtained from the mock catalogue. 
It is compared to the Stripe82X catalogue from \citet{2017ApJ...850...66A} limited to a flux $6\times 10^{-15}$. 
We find a good agreement. 
Based on these three tests: optical luminosity function, observed magnitude number density as a function of redshift and logN-logR for X-ray selected AGNs, we deduce that the model is reliable to describe both the X-ray and the optical properties of the AGNs.   

\subsection{Planning spectroscopic follow-up of eROSITA AGN}
\label{subsection:specz:follow:up}

Redshift measurement for eROSITA AGN is key for most scientific applications. 
Ground-based multi-object (typically fiber-fed) spectrographs are suited infrastructures to measure redshifts for large number of sources over wide areas of the sky. 
Based on our mock catalogue predictions (calibrated on observed multi-wavelength data), the optical magnitude range of the eROSITA X-ray selected AGN is broad, thus one needs a set of telescopes to acquire 
spectroscopic redshifts for all AGN in an optimally efficient way. 
1-m class infrastructure can measure redshifts up to magnitude 17-18 \citep[\textit{e.g.} TAIPAN;][]{2017PASA...34...47D}, where cross-talk effects could negatively affect larger aperture telescopes (indeed, SDSS/BOSS spectroscopic surveys had been limited to objects fainter than 17th magnitude); 
2-m class telescope are optimally suited to follow up systems up to 20-21 magnitude \citep[\textit{e.g.} SDSS; ][]{2012AJ....144..144B}, 
while 4-m class telescope could measure redshift up to magnitude 22-23 \citep[\textit{e.g.} 4MOST, DESI; ][]{2012SPIE.8446E..0TD,DESI2016},
and 8-m class telescopes up to 25-26 \citep[\textit{e.g.} VIMOS, MUSE, MSE ][]{2005PASP..117.1284S,2015AA...575A..75B,2016arXiv160600043M}. 

In order to determine an ideal strategy for the systematic spectroscopic follow-up of eROSITA AGN, we first divide the full sky in two regions: the ecliptic pole regions, containing AGN with ecliptic latitude above 80 or below -80 degrees, which are deeply exposed by the SRG/eROSITA survey scanning strategy, and the extra-Galactic region (`Xgal') defined by a Galactic latitude cut at $|b_{gal}|>20^\circ$ but where the poles are removed. 
Using the medium resolution optical templates, the SDSS $r$-band magnitude and the 4MOST exposure time calculator \citep[][version of December 2019]{2016SPIE.9910E..1QD}, we estimate the exposure time needed (expressed in fiber-hours) for each source to reach a median signal-to-noise ratio of 2.5 per Angstrom, which should yield a success rate in measuring redshifts above 95\%. 
For simplicity, we then scaled this to a corresponding exposure time with a 1, 2 and 8-meter class telescopes by dividing or multiplying exposure times by the ratio of the collecting area. 
Table \ref{tab:AGN:spectroscopy}, (\ref{tab:AGN:spectroscopy:eRASS3}, \ref{tab:AGN:spectroscopy:SNR3}) shows the results of the ETC calculation for the eRASS8 (eRASS3, SNR3) scenario. 
About 2.3 million fiber-hours split between these infrastructures would be sufficient to complete the full spectroscopic identification of all eRASS8 AGN. 
Table \ref{tab:AGN:spectroscopy}, (\ref{tab:AGN:spectroscopy:eRASS3}, \ref{tab:AGN:spectroscopy:SNR3}) also shows the average fiber-hours density needed, i.e. the product of the average number of targets per unit area times the average exposure time (in hours). 
In the extra-Galactic area, the average density of fiber-hours required is on the low side: 20-50/deg$^2$. 
In the deep ecliptic pole regions, the density of fiber-hours required is higher: 100 to 400 /deg$^2$. 
The fiber-hour densities required are within reach of current or near-future state-of-the art multi-objects spectrograph instruments, such as: 
\begin{itemize}
 \item 1m-class: TAIPAN (25 fibers/deg$^2$, 150 fibers),
 \item 2m-class: SDSS  (145 fiber/deg$^2$, 1000 fibers),
 \item 4m-class: 4MOST (400 fibers/deg$^2$, 1600 fibers),
 \item 8m-class: MOONS (7200 fiber/deg$^2$, 1000 fibers).
\end{itemize}
The fiber-hour densities required are well below availability, so the eROSITA AGN follow-up can be carried out next to other programs. 
At the time of writing, large spectroscopic follow-up campaigns of the eROSITA AGN population are planned for SDSS-V \citep{Kollmeier17} and 4MOST \citep{2012SPIE.8446E..0TD,2019Msngr.175...42M} 

\begin{table}
	\centering
	\caption{Number of fiber-hours needed to follow-up spectroscopically all eROSITA AGN in the optical band. 
	D: diameter of the telescope in meters. 
	magnitude: magnitude selection range. 
	fiber-hours: total number of fiber-hours to cover the extra-Galactic area, or the poles and number density of fiber-hours.}
	\label{tab:AGN:spectroscopy}
	\begin{tabular}{ccc rr } 
	\hline   \hline
D   & \multicolumn{2}{c}{magnitude} & \multicolumn{2}{c}{fiber-hours} \\ 
m &  min & max               & total & density [deg$^{-2}$] \\     \hline
\multicolumn{5}{c}{eRASS8 Xgal} \\
\hline
1m & 12.0 & 18.0 &    4,470.67 & 0.17 \\
2m & 18.0 & 21.0 &  566,463.87 & 21.34 \\
4m & 21.0 & 22.8 &1,094,538.60 & 41.24 \\
8m & 22.8 & 25.0 &  406,647.20 & 15.32 \\
\hline
\multicolumn{5}{c}{eRASS8 poles} \\ \hline
1m & 12.0 & 18.5 &      374.40 & 0.62 \\
2m & 18.5 & 21.5 &   67,921.70 & 113.2 \\
4m & 21.5 & 23.0 &  156,087.72 & 260.15 \\
8m & 23.0 & 26.0 &  112,546.52 & 187.58 \\
\hline  \hline
	\end{tabular}
\end{table}

\subsection{Expected uncertainty in the X-ray luminosity function}
\label{subsection:XLF:prediction}
\begin{figure}
\centering
\includegraphics[width=8cm]{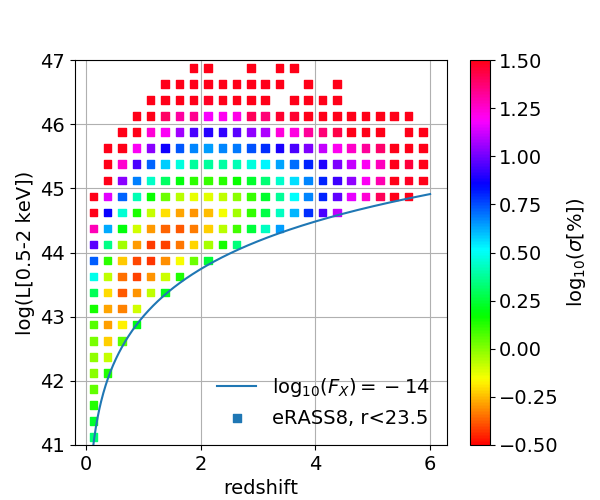}
\caption{Fractional error ($\log_{10}(N^{-0.5})$) on number density measurements in bins of redshift and soft X-ray luminosity (0.5-2 keV band) for soft fluxes brighter than $10^{-14}\mathrm{erg cm}^{-2} \mathrm{s}^{-1}$ in the 0.5-2 keV band (the typical eROSITA flux limit). }
\label{fig:XLF:error}
\end{figure}

Assuming that we will have measured spectroscopically 90\% of the redshifts of all AGN with magnitude $r<23.5$, we predict the expected fractional uncertainty in the luminosity function, see Fig \ref{fig:XLF:error}. 
At low redshift, percent precision will be reached over a wide range of luminosities and will lead to a strong anchor of the XLF models. 
At intermediate redshift $1<z<3$, the uncertainty will be around the percent level at luminosities of $logL_X~44-45$ (i.e. just above the knee of the XLF), an improvement of more than one order of magnitude with respect to previous studies (which constrained it at the 20-50\% level) \citep{Miyaji2015ApJ804104M,2015MNRAS.451.1892A}. 
This will strongly constrain the evolution of the bright end of the XLF. 

\subsection{Clustering on large scales, and the baryon acoustic oscillation standard ruler}
\label{subsection:clustering}
The upcoming combination of eROSITA + SDSS-V + 4MOST will increase the size of the current largest collection of X-ray selected AGN with spectroscopic redshift measurements by a factor of 100. This will allow precise clustering measurements as a function of redshift. 
eROSITA is unique in terms of target selection as it spans a very wide redshift range.

Based on the redshift distribution, we predict the quality of clustering measurements on large scales and the precision with which one could measure the baryon acoustic oscillation standard ruler \citep{eisenstein_98,Seo_2003,Seo2007ApJ...665...14S}. 
Such a calculation requires two more inputs as a function of redshift: the bias and the redshift uncertainty. 
We measure the clustering amplitude of X-ray AGN in the mock catalogue as a function of redshift and find that the description of the bias evolution with redshift is in agreement with Eq. (12) of \citet{Chehade2016MNRAS4591179C}. 
We assume redshift uncertainties to be constant $\sigma_z=0.005(1+z)$, which is slightly more conservative than what is found in \citet{2015ApJS..221...27M,sergio_2017MNRAS468728R}. 

\begin{figure*}
\centering
\includegraphics[width=15cm]{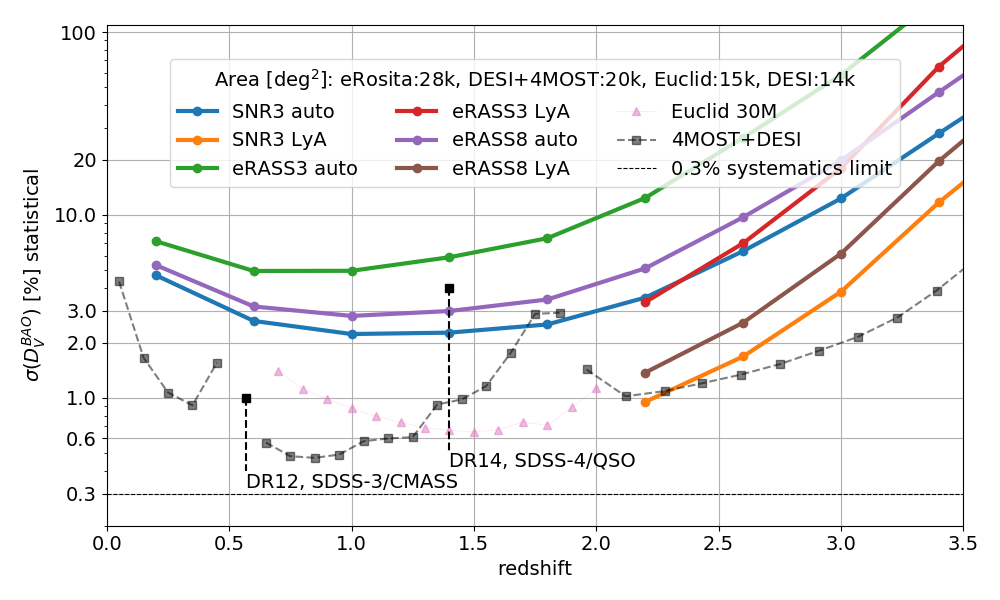}
\caption{Expected statistical uncertainty on the angle averaged BAO scale for 4 surveys: eROSITA, DESI, Euclid, DESI+4MOST. The areas covered in square degrees area: eROSITA: 28k, DESI: 14k, 4MOST: 6k (DESI+4MOST: 20k), Euclid: 15k (30e6 galaxies scenario). For eROSITA, we consider all eRASS3, eRASS8 and SNR3 scenarios.}
\label{fig:BAO:forecast}
\end{figure*}

Applying the Fisher formalism \citep{2014JCAP...05..023F,2016MNRAS.457.2377Z}, we find that the SNR3 sample (over 28k deg$^2$ i.e. the full extra-Galactic sky) will provide a BAO measurement better than 3\% from redshift 0.5 to 2 using the AGN auto-correlation and the same accuracy from redshift 2 to 3 using the Ly$\alpha$ forest (see Fig. \ref{fig:BAO:forecast}), in good agreement with \citet{2014AA572A28H}. 
Dedicated BAO/clustering experiments such as DESI, 4MOST and Euclid will of course provide better formal constraints, 
but eROSITA has the advantage that it will be based on a simple X-ray target selection. This may reduce systematics in the measurements of the clustering on large scales, or at minimum provide a cross-check to be compared to other clustering measurements. 

\subsection{Clustering on small scales: the constraining power of eROSITA}
\label{subsec:ssc}

Exploring clustering as a function of various AGN parameter receives a large interest in the community. 
\citet{2017A&A...600A..97R} studied the variation in the AGN clustering signal as a function of radio-loudness; 
\citet{2013ApJ...778...98S,2018PASJ...70S..33H} \textit{v.s.} optical luminosity; 
\citet{2014ApJ...796....4A,2016ApJ...832..111J,2018MNRAS.474.1773K} \textit{v.s.} AGN type; 
\citet{2016MNRAS.456..924D} \textit{v.s.} obscuration; 
\citet{2010ApJ...713..558K,2013MNRAS.428.1382K,2016MNRAS.457.4195M}  \textit{v.s.} X-ray luminosity; 
\citet{2016ApJ...821...55M} \textit{v.s.} the wavelength AGNs were selected in IR or X-ray or radio; 
\citet{2009ApJ...697.1656S,2013ApJ...775...43K,2015ApJ...815...21K} as a function of black hole mass, 
\citet{2015ApJ...815...21K} \textit{v.s.} Eddington ratio.  
Among others, \citet{2016MNRAS.456..924D} found a hint of a discrepancy between the clustering of obscured vs. unobscured AGNs at redshift 1 (one large redshift bin 0$<z<$2 over $\sim$3000 deg$^2$). 
They find obscured AGNs to have a bias of $b=2.13\pm0.14$ and unobscured of $b=1.82\pm0.11$. 
Uncertainties ($\sim 0.1-0.15$) are large due to the use of angular clustering integrated over a large redshift range. 

\begin{figure}
\centering
\includegraphics[width=8cm,type=png,ext=.png,read=.png]{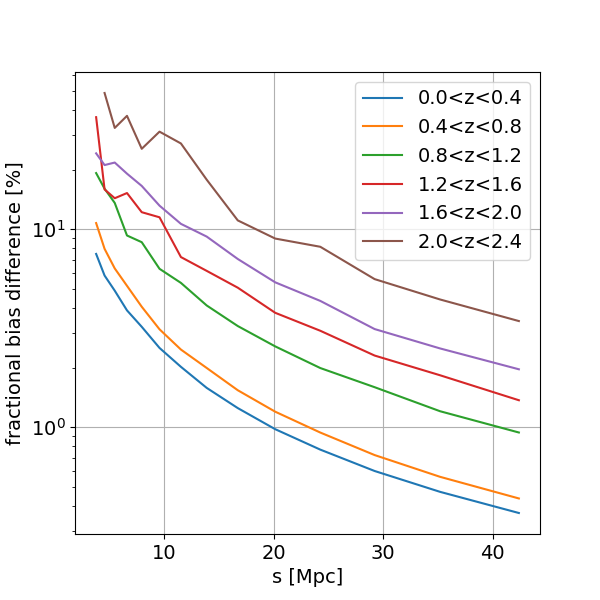}
\caption{Sensitivity to changes in the clustering amplitude (bias) as a function of scale. Only Poisson noise is considered.}
\label{fig:bias:err}
\end{figure}

We predict the eROSITA performance relative to the \citet{2016MNRAS.456..924D} measurements. 
We split the AGN population into optically obscured and optically unobscured sources in the eRASS8 mock catalogue over half of the extra-Galactic sky and in redshift bins of width $dz=0.4$ (considering only sources with $r<23.5$ i.e. those that will have reliable spectroscopic redshifts). 
We count pairs with separations between 3.4 and 46.7 Mpc in 15 logarithmic bins of separation. 
Assuming an uncertainty dominated by Poisson noise, we evaluate the fractional difference in the correlation function that the upcoming eRSS8 AGN sample will be sensitive to, as a function of scale and redshift. 
We find that at low redshift, where both obscured and unobscured AGNs are present in large numbers, the eRASS8 data will be sensitive to percent level variations of the clustering amplitude at separations greater than 20 Mpc
(see Fig. \ref{fig:bias:err}). 
This implies that we should easily be able to measure the suggested deviation in \citet{2016MNRAS.456..924D} at high significance. 
We will furthermore be able to measure accurate differences as a function of redshift and make evolutionary statements. 
The current mock does not have a satellite model, we thus cannot predict the clustering at scales smaller than 5 Mpc. 
At higher redshift, the density of AGNs drops so, for example, we will only be able distinguish deviations at the 10\% level from one another at redshift 2. 

\section{Discussion and conclusions}
\label{sec:conclusion}

\subsection{Enhancing the eROSITA science with end to end simulations}
\label{sec:end2end}

In this section we discuss how the mock should help to interpret the eROSITA observations. 
The mock catalogue is a test bench to validate algorithms either in a forward modelling approach (convolution of observational biases) or in a backward modelling fashion (deconvolution) \citep[\textit{e.g.}][]{ross2017MNRAS4641168R,2018PhRvD..98d2006E,2018arXiv180405867Z}. 
Here we walk through the forward model approach that we envision using to create the eROSITA AGN large scale structure catalog. 

The first step is to simulate the photon events corresponding to the AGN catalogue. 
Then feed event files in the general eROSITA pipeline to create an `observed' AGN catalog. 
This will contain positions, fluxes and detection significance. 
The direct comparison of the input mock and the observed catalogue will allow further refinement of the sensitivity curves obtained by \citet{2018A&A...617A..92C}. 
In particular, we will measure and tabulate the sensitivity curves and their variations with right ascension, declination, exposure time and detection significance. 
Additionally, the comparison of the observed catalogue with the input catalogue will enable us to measure position uncertainties relative to true positions, and biases in the recovered fluxes as a function of AGN type, redshift and luminosity. 
It will also fully justify the detection probability thresholds to be used as a function of the completeness and purity desired in any given sample. 
We will study how these quantities converge as a function of the total exposure time and the number of passes (i.e. from eRASS1 to eRASS8). 
The mock catalogue will also allow quantification of the impact of source confusions on the construction of the catalogue at the poles, where the exposure is deepest, accounting properly for clustering of the sources. 

A second important application is to test the optical infrared counterpart finding algorithm \citep{2018MNRAS.473.4937S} and the photometric redshift estimates \citep{2011ApJ...742...61S}. 
For the first step, having an accurate X-ray position uncertainty is very important, but the counterpart identification process can also be tested explicitly by adding simulated optical counterparts of mock X-ray sources to real optical identification catalogs. 
This will allow an assessment the convergence, purity and completeness of the counterpart finding algorithm as a function of 
\begin{itemize}
    \item the X-ray parameters determined in the previous step;
    \item depth of the optical and IR surveys: the depth of the current realization is tuned for the combination of HSC, VHS and WISE;
    \item the environment: density of galaxies, type of galaxies. 
\end{itemize}
Once the counterparts to the synthetic sources are identified, their photometric redshifts can be estimated using the standard tools and compared with the known redshift from the mock catalogue. This will give a robust estimate of the photometric redshift error and the fraction of catastrophic outliers without the need for extensive spectroscopy and without limitation in terms of object numbers. 
Current estimates suggest that the fraction of catastrophic outliers may be as large as 15\% with the assumed bands. 
The mock catalogue will allow the creation of sub samples with the most reliable photometric redshifts and/or refinement of the methods. It should also be possible to test whether studies of clustering with photometric redshifts can give reliable results. 

A further application is the validation of the measured (photometric or spectroscopic) redshift distributions as a function of the position on the sky and the relative depths of the multi-wavelength surveys used. 
This will permit the quantifications of the completeness achieved as a function of redshift and position on the sky. 
Because the mock is able reproduce the input luminosity function, we should derive completeness weights with good accuracy. 
These then feed into the measurements of the luminosity function from the real data, allowing the a measurement of the luminosity function exploting the full statistical power of the data, uninhibited by completeness effects. 

We will also use the mock to validate angular clustering measurement as a function of X-ray luminosity thresholds \textit{e.g.} using volume-limited AGN samples, which are very sparse in current surveys. 
The discovery space around these upcoming clustering measurements is very large. On top of the very precise large-scale halo bias measurement we expect, cross-correlation with other galaxies at smaller scales will contain a vast amount of information and will be a powerful tool to constrain how galaxies co-evolve with AGNs \citep{2005Natur.433..604D, 2006ApJS..163....1H, 2008ApJS..175..356H, 2008MNRAS.385.1846M, 2011ApJ...741L..33B, Fanidakis2011MNRAS41053F, 2013MNRAS.435..679F, 2014ARA&A..52..589H}. 
To measure the clustering, we need a set of random points that are imprinted with the sensitivity map and systematic biases induced by the two steps described above.
To do so we will run the same analysis as described above except that right ascension and declination will be randomly distributed on the celestial sphere. 
In this manner we obtain a set of random points that by construction contains all the possible systematics. 
The forward modelling approach allows clustering measurements down to any angular separation while the backward modelling approach has a minimum scale below which clustering cannot be recovered. 
Additional milky way, stellar, dust masks and spectroscopic follow-up selection function that add a weight to entire regions of the sky are applied a posteriori on both data and random positions. 

In summary this forward model approach guarantees an optimal control over possible systematic errors affecting the estimation of the number density as a function of sky position, redshift, and depth (X-ray or optical and IR). 

\subsection{On small scale clustering}
\label{subsubsec:hod:guo}
While there is clear observational evidence that super massive black holes and galaxies evolve together, the physical reason remains unclear \citep{2000ApJ...539L..13G,2000ApJ...539L...9F,2015AARv..23....1B}. 
There are two main paths of co-evolution. 
In one, the AGN evolves jointly with its galaxy host, with its accretion rate to some extent correlated to star formation rate \citep{2005Natur.433..604D, 2006ApJS..163....1H, 2008ApJS..175..356H, 2008MNRAS.385.1846M}. 
In the other, AGNs would occur randomly in the galaxy population with some duty cycle \citep{2011ApJ...741L..33B, Bo2016AA588A78B}, with some kind of self-regulatory mechanism imprinting the BH-galaxy correlation.
In the future we intend to extend our model to create predictions for the different scenarios of AGN-host galaxy coevolution. 
We will use semi-analytical models of the galaxy population containing central and satellite galaxies \citep{2016MNRAS.462.3854L}. 
Our empirical AGN X-ray model then shows us how to add X-ray AGNs to the SAM model and study AGN relative to other galaxy populations \citep[\textit{e.g.}][]{2009MNRAS.396..423B} which can subsequently be searched for and tested. 

\subsection{Summary and conclusion}
We have described a new methodology to construct mock catalogues based on dark matter only N-body simulations for X-ray selected AGN samples. 
By construction, the model reproduces accurately ($\pm5\%$) the 2-10 keV luminosity functions up to redshift 6 as well as the observed AGN logN-logS function (to within $\pm20\%$). 
We create an empirical AGN obscuration model that jointly classifies AGNs in the optical and the X-ray. 
From the observed distribution of AGNs in the X-ray flux -- optical $r$-band magnitude plane, we predict optical properties for AGNs, obtaining reasonable agreement with the most recent published optical AGN luminosity functions \citep{Ross_N_2013ApJ77314R,Palanque2016AA587A41P,Caditz2017AA608A64C}. 
Finally we tabulate a set of X-ray spectra, UV-optical-IR SEDs and optical medium resolution spectra that represent well the AGN population present in the mock catalogue. 

This is used to predict the redshift distribution of the AGN to be detected by eROSITA, showing good agreement with previous estimates \citet{arxiv12093114_Merloni,Hutsi2014AA561A58H}.
We show that the XLF will be recovered to an extremely high level of accuracy, more than a factor of 10 better than current state-of-the-art estimates. 
Our model also predicts the large-scale three dimensional distribution of the eROSITA AGN. The large scale bias of halo hosting AGNs is in agreement with previous estimates \citep{2015ApJ...815...21K,2018MNRAS.474.1773K,Chehade2016MNRAS4591179C,sergio_2017MNRAS468728R}. Once again the eROSITA survey will yield clustering measurements of X-ray selected AGN with unprecedented accuracy. Finally we predict the measurements expected of the BAO standard ruler and find it will be a very interesting data set for large scale structure studies, complementing dedicated cosmological experiments, but with arguably a simple selection function. 

The mock catalogue is a useful tool for full and accurate science exploitation of the eROSITA all-sky survey: It 1) provides accurate forecasts and estimates of the power of the eROSITA survey, 2) can validate analysis methods, 3) estimate a wide variety of errors, biases, incompletenesses and correction factors.

\section*{Acknowledgements}
J. Comparat thanks M. Freyberg, K. Dennerl, S. Bonoli, H. Guo, J. Aird, T. Miyaji for fruitful discussions about this project. 

JB acknowledges support from the CONICYT-Chile grants Basal-CATA PFB-06/2007 \& AFB-170002, FONDECYT Postdoctorados 3160439 and the Ministry of Economy, Development, and Tourism's Millennium Science Initiative through grant IC120009, awarded to The Millennium Institute of Astrophysics, MAS. This research was supported by the DFG cluster of excellence “Origin and Structure of the Universe”.

GY would like to thank MINECO/FEDER for financial support under project grant AYA2015-63810-P.

M.K. acknowledges support from DLR grant 50OR1802.

The CosmoSim database is a service of the Leibniz-Institute for Astrophysics Potsdam (AIP).
The MultiDark database was developed in cooperation with the Spanish MultiDark Consolider Project CSD2009-00064.

\bibliographystyle{mnras}
\bibliography{references}

\appendix

\section{Tables}

\begin{table*}
	\centering
	\caption{\label{tab:AGN:NZ:ALL:eRASS3}Same as Table \ref{tab:AGN:NZ:ALL:eRASS8} for the eRASS3 scenario.}	
	
	\begin{tabular}{cc rr rrr rrr rrrrrrrrrrr} 
	\hline   
\hline
\multicolumn{2}{c}{$z$ bin}  &
\multicolumn{2}{c}{$12<r<18$} & 
\multicolumn{2}{c}{$18<r<21$} & 
\multicolumn{2}{c}{$21<r<22.8$} & 
\multicolumn{2}{c}{$22.8<r<25$} &
\multicolumn{2}{c}{type 1 $r<25$} &
\multicolumn{2}{c}{all} \\ 

min & max &
 x & p &
 x & p &
 x & p &
 x & p &
 x & p &
 x & p \\
\hline
     0.0 &     0.4 &     0.3075 &     0.2569 &     2.9120 &     4.6591 &     0.8473 &     3.5422 &     0.0620 &     0.6095 &     1.9963 &     3.5917 &     4.1294 &     9.0901 \\ 
     0.4 &     0.8 &     0.3156 &     0.3255 &     6.1014 &    11.2649 &     2.2519 &    10.0411 &     0.1736 &     1.8573 &     7.3422 &    15.9719 &     8.8436 &    23.5302 \\ 
     0.8 &     1.2 &     0.2215 &     0.2122 &     5.9180 &    12.1999 &     2.6236 &    12.1999 &     0.2243 &     2.3982 &     8.2504 &    21.7751 &     8.9892 &    27.0677 \\ 
     1.2 &     1.6 &     0.1268 &     0.1436 &     4.3450 &    10.2916 &     2.2321 &    11.2777 &     0.2090 &     2.4620 &     6.3860 &    20.3582 &     6.9142 &    24.2227 \\ 
     1.6 &     2.0 &     0.0711 &     0.0798 &     2.8172 &     7.7402 &     1.6417 &     9.6055 &     0.1751 &     2.1812 &     4.3386 &    16.6117 &     4.7065 &    19.6545 \\ 
     2.0 &     2.4 &     0.0309 &     0.0399 &     1.3617 &     4.2028 &     0.9154 &     6.1271 &     0.1093 &     1.4759 &     2.2114 &    10.1767 &     2.4182 &    11.8744 \\ 
     2.4 &     2.8 &     0.0098 &     0.0112 &     0.5110 &     1.8684 &     0.3786 &     2.9152 &     0.0506 &     0.7818 &     0.8590 &     4.7692 &     0.9503 &     5.5894 \\ 
     2.8 &     3.2 &     0.0029 &     0.0000 &     0.1834 &     0.7068 &     0.1511 &     1.3563 &     0.0242 &     0.4037 &     0.3236 &     2.1604 &     0.3621 &     2.4827 \\ 
     3.2 &     3.6 &     0.0010 &     0.0016 &     0.0569 &     0.2409 &     0.0527 &     0.4867 &     0.0096 &     0.1915 &     0.1054 &     0.7739 &     0.1204 &     0.9270 \\ 
     3.6 &     4.0 &     0.0003 &     0.0000 &     0.0193 &     0.0654 &     0.0188 &     0.1628 &     0.0036 &     0.0750 &     0.0363 &     0.2537 &     0.0420 &     0.3032 \\ 
     4.0 &     4.4 &     0.0001 &     0.0016 &     0.0067 &     0.0367 &     0.0064 &     0.0702 &     0.0014 &     0.0271 &     0.0122 &     0.1149 &     0.0146 &     0.1356 \\ 
     4.4 &     4.8 &     0.0001 &     0.0000 &     0.0025 &     0.0160 &     0.0028 &     0.0335 &     0.0006 &     0.0080 &     0.0048 &     0.0431 &     0.0059 &     0.0574 \\ 
     4.8 &     5.2 &     0.0000 &     0.0000 &     0.0010 &     0.0112 &     0.0011 &     0.0112 &     0.0001 &     0.0016 &     0.0017 &     0.0176 &     0.0022 &     0.0239 \\ 
     5.2 &     5.6 &     0.0000 &     0.0000 &     0.0003 &     0.0016 &     0.0004 &     0.0048 &     0.0001 &     0.0048 &     0.0006 &     0.0080 &     0.0009 &     0.0112 \\ 
     5.6 &     6.0 &     0.0000 &     0.0000 &     0.0001 &     0.0000 &     0.0001 &     0.0000 &     0.0000 &     0.0000 &     0.0002 &     0.0000 &     0.0003 &     0.0000 \\ \hline
     0.0 &     6.0 &     1.0875 &     1.0722 &    24.2363 &    53.3057 &    11.1241 &    57.8340 &     1.0435 &    12.4776 &    31.8688 &    96.6261 &    37.4997 &   124.9702 \\ 
\hline 
\end{tabular}
\end{table*}

\begin{table*}
	\centering
	\caption{\label{tab:AGN:NZ:ALL:SNR3}Same as Table \ref{tab:AGN:NZ:ALL:eRASS3} for the SNR3 scenario.}	
	\begin{tabular}{cc rr rrr rrr rrrrrrrrrrr} 
	\hline   \hline
\multicolumn{2}{c}{$z$ bin}  &
\multicolumn{2}{c}{$12<r<18$} & 
\multicolumn{2}{c}{$18<r<21$} & 
\multicolumn{2}{c}{$21<r<22.8$} & 
\multicolumn{2}{c}{$22.8<r<25$} &
\multicolumn{2}{c}{type 1 $r<25$} &
\multicolumn{2}{c}{all} \\ 

min & max &
 x & p &
 x & p &
 x & p &
 x & p &
 x & p &
 x & p \\ \hline
     0.0 &     0.4 &     0.3166 &     0.2585 &     4.4496 &     5.6899 &     2.6049 &     7.4881 &     0.3889 &     2.6966 &     3.1866 &     5.2399 &     7.7644 &    16.2304 \\ 
     0.4 &     0.8 &     0.3417 &     0.3271 &    10.4147 &    14.3077 &     7.3314 &    22.2442 &     1.1525 &     8.4950 &    13.8005 &    23.7393 &    19.2528 &    45.6548 \\ 
     0.8 &     1.2 &     0.2490 &     0.2170 &    11.3186 &    16.4027 &     9.2570 &    27.7235 &     1.5187 &    10.9857 &    18.6644 &    37.4024 &    22.3592 &    55.6719 \\ 
     1.2 &     1.6 &     0.1496 &     0.1468 &     9.2547 &    14.3460 &     8.6079 &    26.7023 &     1.4894 &    11.1484 &    16.7610 &    38.1826 &    19.5168 &    52.6786 \\ 
     1.6 &     2.0 &     0.0861 &     0.0814 &     6.6144 &    11.3766 &     6.9738 &    23.4904 &     1.2873 &    10.0235 &    12.9902 &    33.8202 &    14.9769 &    45.2974 \\ 
     2.0 &     2.4 &     0.0377 &     0.0399 &     3.5655 &     6.7111 &     4.3219 &    16.4378 &     0.9012 &     7.6381 &     7.6889 &    23.8079 &     8.8379 &    31.0822 \\ 
     2.4 &     2.8 &     0.0128 &     0.0112 &     1.4917 &     3.1752 &     2.0872 &     9.1252 &     0.4947 &     4.5474 &     3.5570 &    13.3041 &     4.0930 &    17.0202 \\ 
     2.8 &     3.2 &     0.0036 &     0.0000 &     0.5752 &     1.3802 &     0.9150 &     4.5203 &     0.2495 &     2.6918 &     1.5036 &     6.8132 &     1.7479 &     8.7072 \\ 
     3.2 &     3.6 &     0.0015 &     0.0016 &     0.1931 &     0.4898 &     0.3479 &     1.9977 &     0.1090 &     1.4025 &     0.5590 &     3.0572 &     0.6538 &     3.9491 \\ 
     3.6 &     4.0 &     0.0003 &     0.0000 &     0.0692 &     0.1755 &     0.1300 &     0.8153 &     0.0461 &     0.6159 &     0.2087 &     1.2749 &     0.2465 &     1.6387 \\ 
     4.0 &     4.4 &     0.0002 &     0.0016 &     0.0260 &     0.0830 &     0.0494 &     0.3143 &     0.0197 &     0.2856 &     0.0792 &     0.5409 &     0.0956 &     0.6909 \\ 
     4.4 &     4.8 &     0.0001 &     0.0000 &     0.0097 &     0.0335 &     0.0194 &     0.1532 &     0.0078 &     0.1340 &     0.0308 &     0.2441 &     0.0371 &     0.3271 \\ 
     4.8 &     5.2 &     0.0000 &     0.0000 &     0.0037 &     0.0223 &     0.0084 &     0.0670 &     0.0032 &     0.0543 &     0.0126 &     0.1069 &     0.0153 &     0.1436 \\ 
     5.2 &     5.6 &     0.0000 &     0.0000 &     0.0014 &     0.0032 &     0.0036 &     0.0319 &     0.0013 &     0.0176 &     0.0051 &     0.0399 &     0.0063 &     0.0543 \\ 
     5.6 &     6.0 &     0.0000 &     0.0000 &     0.0006 &     0.0000 &     0.0017 &     0.0128 &     0.0011 &     0.0160 &     0.0027 &     0.0223 &     0.0033 &     0.0287 \\ \hline
     0.0 &     6.0 &     1.1991 &     1.0850 &    47.9880 &    74.1968 &    42.6595 &   141.1241 &     7.6703 &    60.7523 &    79.0504 &   187.5958 &    99.6068 &   279.1750 \\ 
  
\hline 
\end{tabular}
\end{table*}

\begin{table}
	\centering
	\caption{\label{tab:AGN:spectroscopy:eRASS3} Same as Table \ref{tab:AGN:spectroscopy} for the eRASS3 scenario}
	\begin{tabular}{ccc rr } 
	\hline   \hline
D   & \multicolumn{2}{c}{magnitude} & \multicolumn{2}{c}{fiber-hours} \\ 
m &  min & max               & total & density [deg$^{-2}$] \\     \hline
\multicolumn{5}{c}{eRASS3 Xgal} \\
\hline
1m & 12.0 & 18.0 &   4,070 & 0.15 \\	
2m & 18.0 & 21.0 & 312,051 & 11.76 \\
4m & 21.0 & 22.8 & 372,991 & 14.05 \\
8m & 22.8 & 25.0 &  88,220 & 3.32 \\
\hline
\multicolumn{5}{c}{eRASS3 poles} \\ \hline
1m & 12.0 & 18.5 &     367 & 0.61 \\
2m & 18.5 & 21.5 &  45,674 & 76.12 \\
4m & 21.5 & 23.0 &  70,477 & 117.46 \\
8m & 23.0 & 26.0 &  35,201 & 58.67 \\

\\\hline  \hline
	\end{tabular}
\end{table}

\begin{table}
	\centering
	\caption{\label{tab:AGN:spectroscopy:SNR3} Same as Table \ref{tab:AGN:spectroscopy} for the SNR3 scenario}
	\begin{tabular}{ccc rr } 
	\hline   \hline
D   & \multicolumn{2}{c}{magnitude} & \multicolumn{2}{c}{fiber-hours} \\ 
m &  min & max               & total & density [deg$^{-2}$] \\     \hline
\multicolumn{5}{c}{SNR3 Xgal} \\
\hline
1m & 12.0 & 18.0 &    4,569 & 0.17 \\
2m & 18.0 & 21.0 &  711,221 & 26.80 \\
4m & 21.0 & 22.8 &1,717,619 & 64.71 \\
8m & 22.8 & 25.0 &  789,628 & 29.75 \\
\hline
\multicolumn{5}{c}{SNR3 poles} \\ \hline
1m & 12.0 & 18.5 &      375 & 0.63 \\
2m & 18.5 & 21.5 &   78,179 & 130.30 \\
4m & 21.5 & 23.0 &  215,965 & 359.94 \\
8m & 23.0 & 26.0 &  187,819 & 313.03 \\

\hline  \hline
\end{tabular}
\end{table}

\section{Figures}

\begin{figure}
\centering
\includegraphics[width=8cm]{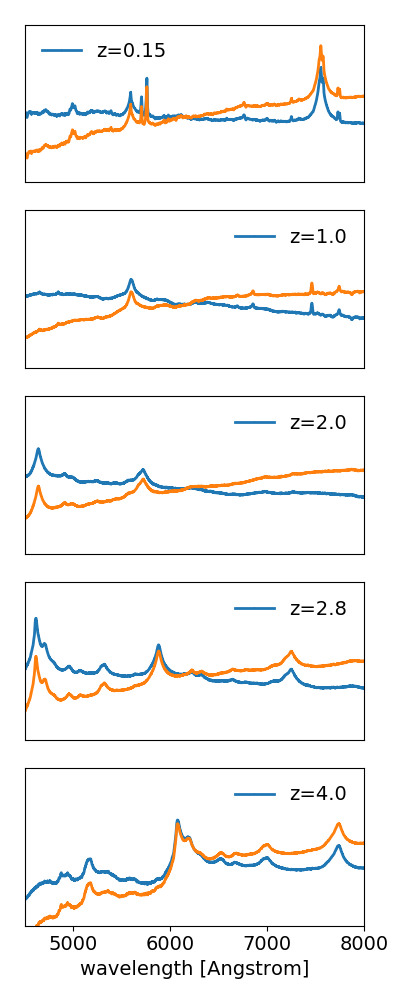}
\caption{Flux (arbitrary units) vs. wavelength of the type 1 spectral templates (blue). The orange line shows the same spectrum redenned with E(B-V)=1. Both are normalzed to an SDSS $r-band$ of 16.5.}
\label{fig:spectra:qso}
\end{figure}

\begin{figure}
\centering
\includegraphics[width=8cm]{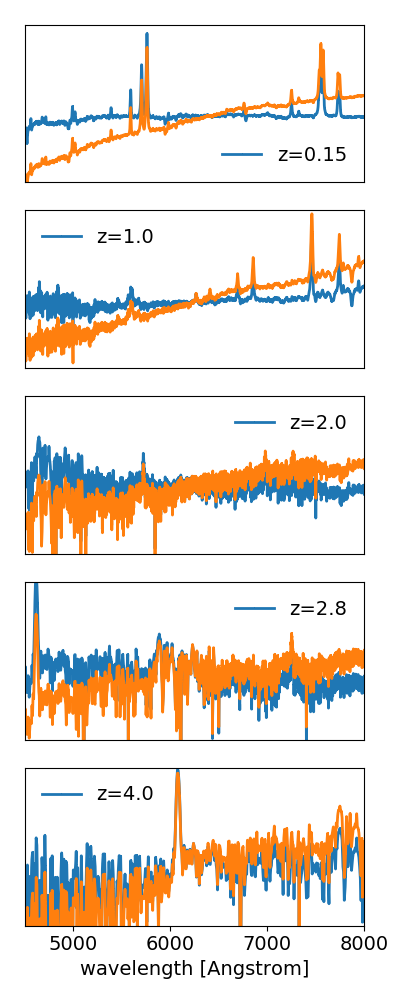}
\caption{Same as \ref{fig:spectra:qso} for the type 2 spectral templates.}
\label{fig:spectra:agn}
\end{figure}

\begin{figure}
\centering
\includegraphics[width=8cm]{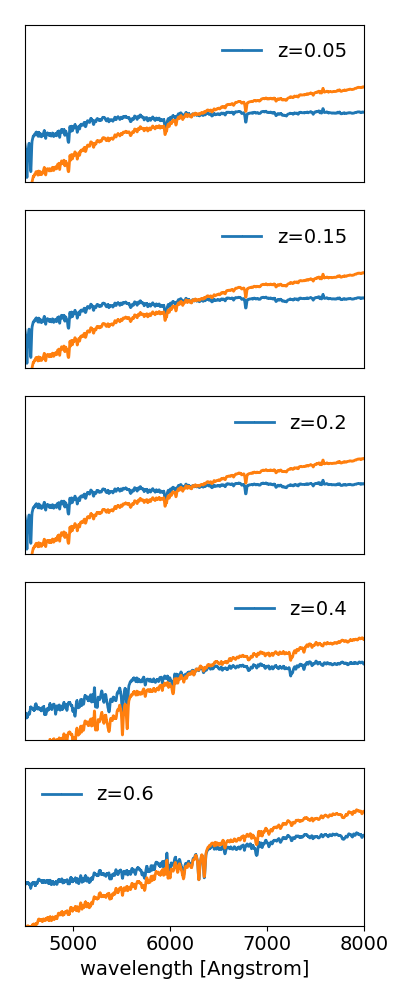}
\caption{Same as \ref{fig:spectra:qso} for the red. elliptical galaxies, used also for the type 3 spectral templates.}
\label{fig:spectra:lrg}
\end{figure}

\begin{figure}
\centering
\includegraphics[width=8cm]{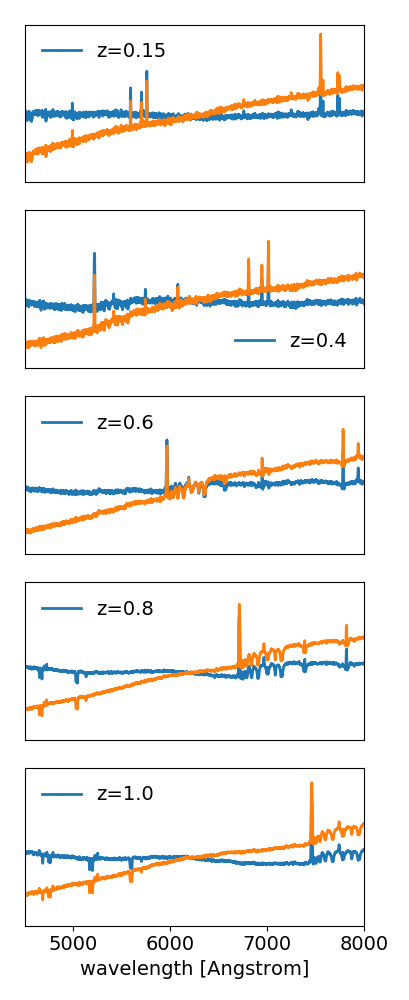}
\caption{Same as \ref{fig:spectra:qso} for the star forming galaxy spectral templates. These are not used in the current mock.}
\label{fig:spectra:elg}
\end{figure}

\begin{figure*}
\centering
\includegraphics[width=17cm]{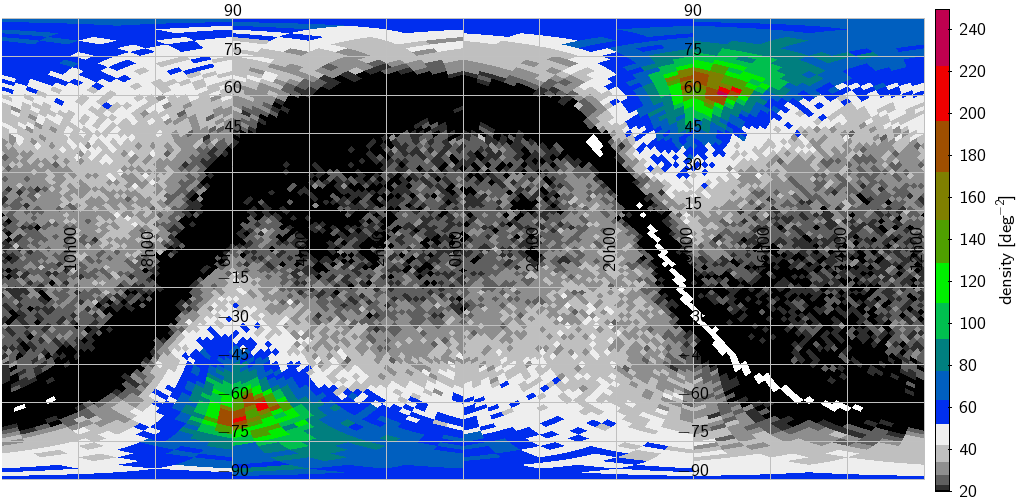}

\includegraphics[width=17cm]{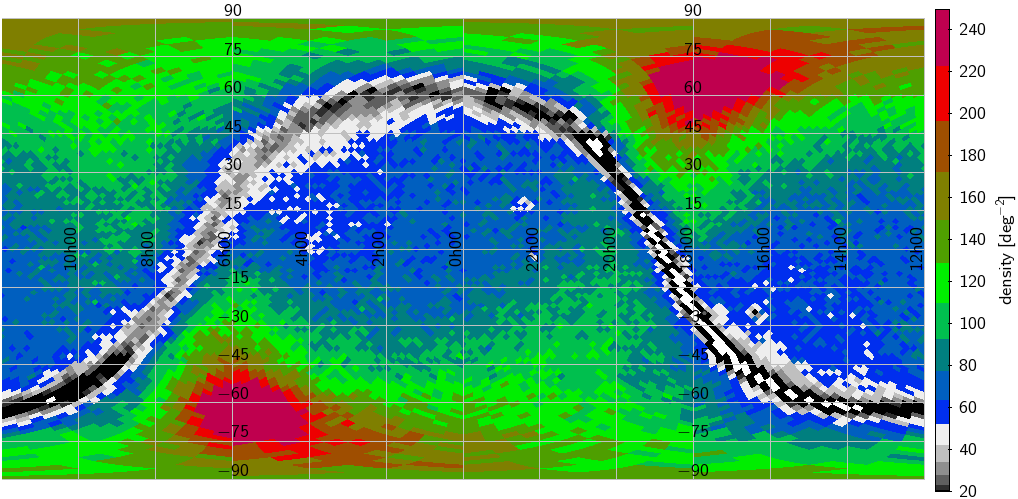}
\caption{Same as Fig. \ref{fig:sky:distributions}. Top panel eRASS3, bottom SNR3.}
\label{fig:sky:distributions:2}
\end{figure*}

\begin{figure}
\centering
\includegraphics[width=\columnwidth,type=png,ext=.png,read=.png]{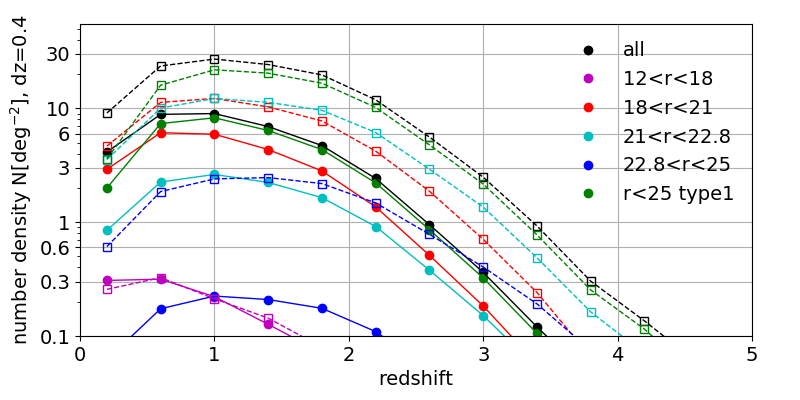}
\caption{\label{fig:redshift:distributions:eRASS3}Same as Fig. \ref{fig:redshift:distributions:eRASS8} for the eRASS3 scenario. 
It corresponds to Table \ref{tab:AGN:NZ:ALL:eRASS3}.
}
\end{figure}

\begin{figure}
\centering
\includegraphics[width=\columnwidth,type=png,ext=.png,read=.png]{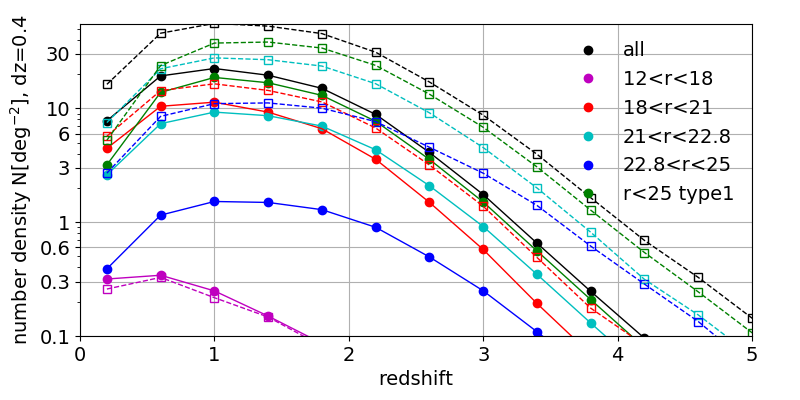}
\caption{\label{fig:redshift:distributions:SNR3}Same as Fig. \ref{fig:redshift:distributions:eRASS8} for the SNR3 scenario. 
It corresponds to Table \ref{tab:AGN:NZ:ALL:SNR3}.
}
\end{figure}

\label{lastpage}
\end{document}